\documentclass[fleqn,usenatbib]{mnras}

\usepackage{newtxtext,newtxmath}
\usepackage[T1]{fontenc}
\usepackage{ae,aecompl}
\usepackage{graphicx}	
\usepackage{amsmath}	
\usepackage[dvipsnames]{xcolor}	
\usepackage{scalefnt}
\usepackage{threeparttable, tablefootnote}
\usepackage{wrapfig}	
\usepackage[textwidth=1.5cm,shadow]{todonotes}	
\usepackage{color, soul}	
\usepackage{verbatim}	
\usepackage{lineno}
\usepackage[normalem]{ulem} 
\usepackage{tcolorbox}  

\newcommand{\orcid}[1]{\href{#1}{\includegraphics[scale=0.035]{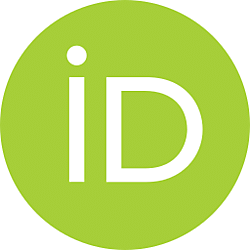}}}

\definecolor{boxback}{HTML}{DAE5F0}
\definecolor{boxframe}{HTML}{0F365B}
{
\end{tcolorbox}   
}

\newcommand{\eg}[1]{(e.g. \citealt{#1})}

\def\code#1{\texttt{#1}}

\title[Simulations of X-ray binaries in the hard state]{Emergence of hot corona and truncated disk in simulations of accreting stellar mass black holes}

\author[R. Nemmen et al.]{
Rodrigo Nemmen$^{1,2}$\thanks{E-mail: rodrigo.nemmen@iag.usp.br}\orcid{https://orcid.org/0000-0003-3956-0331}, 
Artur Vemado$^{1,3}$\orcid{https://orcid.org/0000-0003-3486-4037},
Ivan Almeida$^{1,4}$\orcid{https://orcid.org/0000-0001-6018-2852}, Javier Garcia$^{5,6}$\orcid{https://orcid.org/0000-0003-3828-2448}
\newauthor and Pedro N. Motta$^{1}$
\\
$^{1}$Instituto de Astronomia, Geof\'{\i}sica e Ci\^encias Atmosf\'ericas, Universidade de S\~ao Paulo, S\~ao Paulo, SP 05508-090, Brazil\\
$^2$Kavli Institute for Particle Astrophysics and Cosmology (KIPAC), Stanford University, Stanford, CA 94305, USA \\
$^{3}$LEEGA, S\~ao Paulo, SP 04729-080, Brazil \\
$^4$ School of Mathematics, Statistics and Physics, Newcastle University, NE1 7RU, UK \\
$^5$Cahill Center for Astronomy and Astrophysics, California Institute of Technology, Pasadena, CA 91125, USA \\
$^6$NASA Goddard Space Flight Center, Greenbelt, MD 20771, USA
}

\date{Accepted XXX. Received YYY; in original form ZZZ}

\pubyear{2023}

\begin{document}
\label{firstpage}
\pagerange{\pageref{firstpage}--\pageref{lastpage}}
\maketitle

\begin{abstract}
Stellar mass black holes in X-ray binaries (XRBs) are known to display different states characterized by different spectral and timing properties, understood in the framework of a hot corona coexisting with a thin accretion disk whose inner edge is truncated. There are several open questions related to the nature and properties of the corona, the thin disk, and dynamics behind the hard state. This motivated us to perform two-dimensional hydrodynamical simulations of accretion flows onto a $10 M_\odot$ black hole. We consider a two-temperature plasma, incorporate radiative cooling with bremmstrahlung, synchrotron and comptonization losses and approximate the Schwarzschild spacetime via a pseudo-Newtonian potential. We varied the mass accretion rate in the range $0.02 \leq \dot{M}/\dot{M}_{\rm Edd} \leq 0.35$. Our simulations show the natural emergence of a colder truncated thin disk embedded in a hot corona, as required to explain the hard state of XRBs. We found that as $\dot{M}$ increases, the corona contracts and the inner edge of the thin disk gets closer to the event horizon. At a critical accretion rate $0.02 \leq \dot{M}_{\text {crit }}/\dot{M}_{\rm Edd} \leq 0.06$, the thin disk disappears entirely. We discuss how our simulations compare with XRB observations in the hard state.
\end{abstract}


\section{Introduction}

Stellar mass black holes in binary systems (also known as X-ray binaries, XRBs) are known to display three states characterized by different spectral and timing properties \eg{Remillard2006, Fender2012}. In the low or hard state the source is fainter and its X-ray spectrum displays a hard power-law. As it brightens, it enters a high or soft state, in which the spectrum is characterized by a thermal component. Between these two states the source goes through an intermediate (or very high state) state where the thermal and power-law components are comparable. A black hole walks though all these states in a couple of months, and sometimes days.

It is widely accepted that the hard and thermal states can be understood separately in terms of a hot corona and a thin disk, respectively \citep{Esin1997,Done2007}. The basic qualitative idea is that, in the hard state, the thin disk is truncated at a certain transition radius $R_{\rm tr}$ inside of which it gives way to a hot corona (i.e. a radiatively inefficient accretion flow or RIAF). As the mass accretion rate $\dot{M}$ increases, $R_{\rm tr}$ becomes progressively smaller until the entire accretion flows consists of a thin disk: the source transitions to the soft state. Physically, this is due to the cooling time becoming shorter than the accretion timescale \citep{Narayan1995b}. 

Two unresolved questions in XRB research stand out. The first: what is the nature and properties of the X-ray-emitting corona? The second question: what is the mechanism and dynamics behind state transitions? Observationally, there is firm evidence that both a hot corona and a truncated thin disk are present along the state cycles of XRBs. However, it has been challenging to explain from first principles how a stable corona is formed and its physical properties \eg{Schnittman2013, Yuan2014, Dexter2021}. 
In addition, the details of the transition from the hard to soft state occurs are hotly debated. Is it driven by a decrease in the inner edge of the thin accretion disk \citep{Ingram2011,Plant2014}, or a diminution of the corona \citep{Garcia2015, Kara2019, Wang2021}? This remains unclear. 

What is clear is that radiative cooling in the accretion flow should play a major role in explaining these transitions. While historically analytical one-dimensional models have been very useful in beginning to address these questions \eg{Chen1995, Esin1997}, some aspects of accretion physics such as their dynamics are beyond the scope of such models. 

In this work, we use viscous hydrodynamical simulations designed to mimic the state of the XRB accretion flow in the hard state. We incorporate energy losses due to radiative cooling and investigate a range of mass accretion rates near the critical value $\dot{M} \approx 0.01 \dot{M}_{\rm Edd}$ at which a RIAF collapses to a thin disk. Section \ref{sec:methods} describes the numerical experiment and initial condition of the accretion flow. Section \ref{sec:res}  describes the results of our experiment. Section \ref{sec:disc} discusses the results and how they can help to explain XRB observations. Finally, section \ref{sec:end} summarizes the work. The appendices describe technical details about the methods.

\section{Summary of methods}   \label{sec:methods}

We have adopted the code \code{PLUTO}\footnote{See the data availability section for the code with specific customizations used in this work.} \citep{Mignone2007} to solve the equations of fluid dynamics in spherical coordinates (\citealt{Vemado2020} for the full equations and details). We approximate the Schwarzschild spacetime via the pseudo-Newtonian gravitational potential $\phi = -GM/(r-r_{\text{s}})$ \citep{Paczynsky1980} ($r_s$ is the Schwarzschild radius), as the spin parameter does not seem to be important at first order to understand XRB state transitions \eg{Remillard2006} and we are not addressing jet formation.

We assume an $\alpha$-disk, i.e. an accretion flow with shear stress given by the Shakura-Sunyaev $\alpha$ viscosity  \citep{Shakura1973} where the kinematic viscosity is given by $\nu = \alpha r^{1/2}$ \citep{Landau1959, Stone1999}. Here, $\alpha$ is a constant similar to (but not equal) the standard Shakura-Sunyaev $\alpha$. We set $\alpha = 0.01$ which is a common choice \eg{Wu2016}.

A key aspect of our work are the energy losses via radiative cooling. Below we give a summary of the heating and cooling implementation; we describe it in more details in Appendix \ref{ap:energy}. The radiative cooling rate is given by $Q^{-}_{\rm rad}=Q^{-}_{\rm brem} + Q^{-}_{\rm syn} + Q^{-}_{\rm ssc}$ where we include the relevant processes for stellar-mass BHs: bremsstrahlung emission $Q^{-}_{\rm brem}$ including both the electron-ion and electron-electron interactions, optically thin synchrotron emission $Q^{-}_{\rm syn}$ and comptonized self-synchrotron emission $Q^{-}_{\rm ssc}$. We adopt a series of approximations for the calculation of the cooling rates. 
For the bremsstrahlung emission, we assume that the proton temperature is small enough to neglect its motion and the electrons follow a Maxwellian energy distribution. 
For the synchrotron emission, we adopt the optically thin limit \citep{Pacholczyk1970} and assume that the angle between the velocity vector of the electron and the direction of the local magnetic field follows an isotropic distribution \citep{Narayan1995}. In order to obtain the $Q^{-}_{\rm syn}$, we assume that the observer sees a blackbody source with radius determined by the condition $\nu = \nu_c(r)$ (the self-absorption critical frequency).
We neglect nonlocal radiative transfer effects. For the self-synchrotron Comptonization, we assume Comptonization of synchrotron radiation at the frequency $\nu_c$.
We include the cooling contribution from higher density, optically thick regions of the flow using the cooling rate computed by \cite{Hubeny1990}.

The radiative cooling term $Q^{-}_{\text{rad}}(r, n_e, T_e)$ in the energy equation depends the distance from the BH $r$, the electron temperature $T_e$ and the electron number density $n_e$. We generated a lookup table of $-Q^{-}_{\text{rad}}$ as a function of $n_e$, $r$ and $T_e$ for the following parameter ranges, equally spaced in the log space: $10^{11} \leq n_e/\text{cm}^{-3} \leq 10^{20}$, $2.6 \leq r/r_g \leq 800$ and $10^{6} \leq T_e/{\rm K} \leq 10^{12}$, encompassing the range of typical XRB values \eg{Done2007}. This is much faster than computing cooling rates on-the-fly.

Distances in the simulations are parameterized in terms of the gravitational radius $r_g \equiv GM/c^2=M$. We adopt $M=10M_{\odot}$. 
The grid resolution is $400 \times 400$ elements ($R \times \theta$), logarithmic in $r$ bounded by $2.6 r_g < r < 800 r_g$, linear and non-uniform in $\theta$ with ten cells at $\theta < 15^{\circ}$ $\theta > 165^{\circ}$ and the remaining uniformly distributed, in order to increase resolution towards the equatorial plane. 
The initial condition consists of a stationary torus in dynamical equilibrium \citep{Papaloizou1984} with a constant specific angular momentum. Its starting radius is chosen as $r_{\text{min}}=150 r_g$ and the pressure maximum radius as $r_{0}=200 r_g$, resulting in an initial torus extending all the way to $r_{\text{max}}=300 r_g$. The mass accretion rate is set by the torus maximum density $\rho_0$, which sets its total mass.   

We assume the presence of a local, isotropic, randomly tangled magnetic field characterized by $\beta = P_{\mathrm{gas}}/P_{\mathrm{mag}}=10$ on the assumption that the flow is dominated by gas pressure \eg{Wu2016}.
RIAFs are characterized by a two-temperature plasma \eg{Yuan2014}. We approximate the temperature difference using
\begin{equation}
T_{\mathrm{e}} = \frac{T_i}{2+(r/r_0)^{-1}},
\label{eq:temp}
\end{equation}
where $r_0$ is the radius of maximum density and the ion temperature $T_i$ is taken directly from the simulation. Equation \ref{eq:temp} is based on height-integrated solutions \citep{Narayan1995b,Wu2016}. The effective mean molecular weights of ions and electrons are $\mu_i=1.23$ and $\mu_e=1.14$, corresponding to a hydrogen mass fraction $X=0.75$ \citep{Narayan1995}.

The beginning of each simulation involves a burn-in period during which the radiative cooling is disabled until the flow achieves a quasi-steady state characterized by an almost constant accretion rate at $t \sim 5000 GM/c^3$. More precisely, radiative cooling is turned on at $t=1.4 t_{\text{orb}}$ where $t_{\text{orb}}=2\pi r/v_{\phi} \approx 1.8 \times 10^4 GM/c^3$ is the orbital time of the accretion flow at $r=100 r_s$.

\section{Results}   \label{sec:res}

\begin{figure}
\centering
\includegraphics[width=\linewidth]{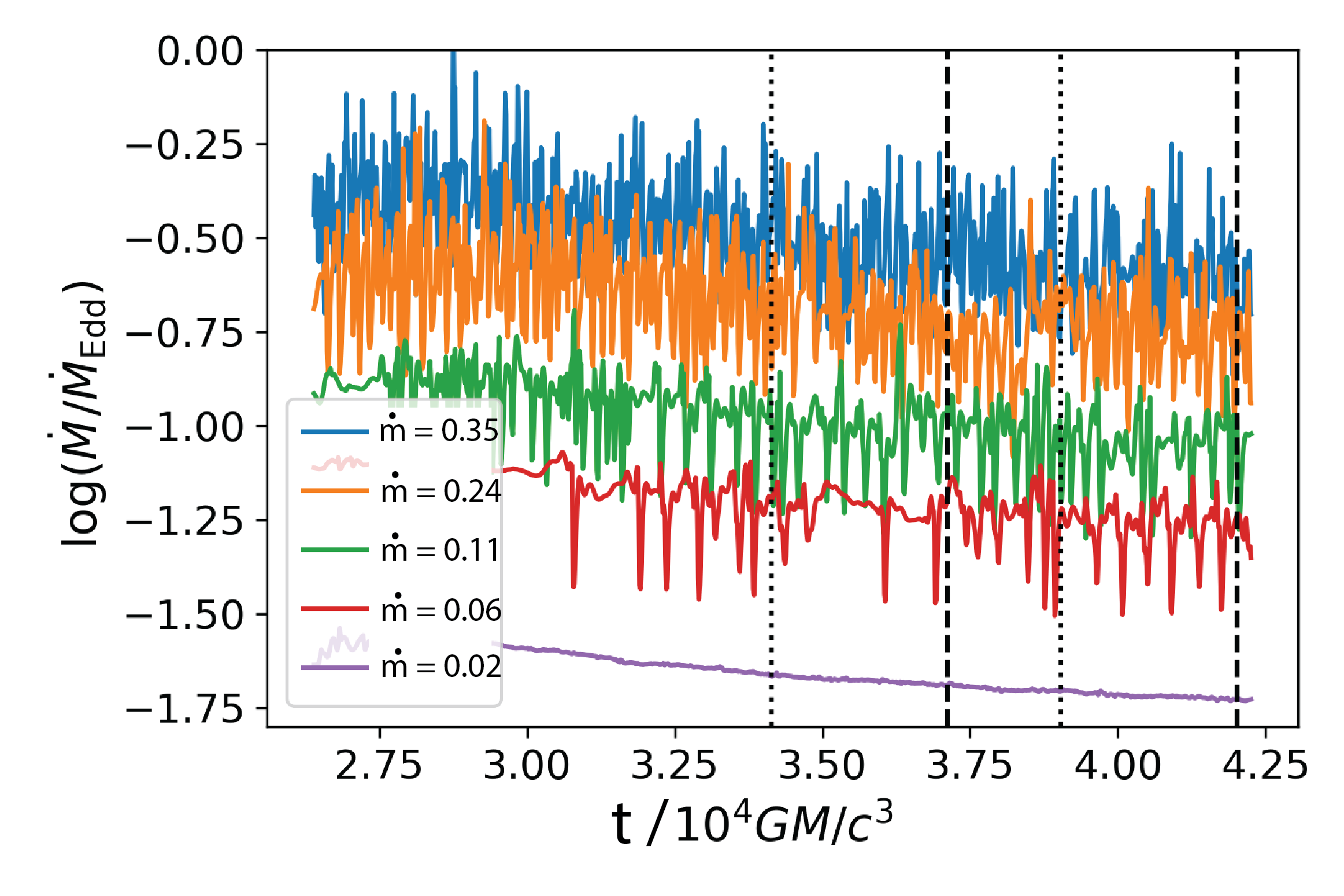}
\caption{Temporal evolution of the mass accretion rate after the radiative cooling is enabled. The dotted vertical indicate the time range used to estimate $\dot{m}$ in models 4 and 5 (cf. Table \ref{tab}); the dashed line shows the range used in the remaining  simulations. } 
\label{fig:mdot-time}
\end{figure}

We carried out five simulations corresponding to different values of the mass accretion rate $\dot{M}$, which are listed in Table \ref{tab}. The mass accretion rates quoted in Table \ref{tab} are the time-average of $\dot{M}(r,t)=2\pi r^2 \int_{0}^{\pi} \rho v_{\mathrm{r}} \sin{\theta}\, d\theta$ at the inner boundary; we computed the average for the time intervals over which $\dot{M}(t)$ is relatively constant thus indicating that the accretion flow reached steady state, as indicated in Figure\ref{fig:mdot-time}. The accretion rates range from $\dot{m}=0.02$ up to $\dot{m}=0.35$, where $\dot{m} \equiv \dot{M}/\dot{M}_{\rm Edd}$, $L_{\rm Edd} \equiv 0.1 \dot{M}_{\rm Edd} c^2$ and $L_{\rm Edd}=1.3 \times 10^{38} (M/M_\odot) \ {\rm erg \ s}^{-1}$. Since the critical value that signals the transition between a RIAF and thin disk is $\dot{m}_{\rm crit} \approx 0.01$ \citep{Yuan2014}, our simulations should probe the conditions involved in the hard to thermal state transition of XRBs. 

\begin{table*}
\centering
\begin{tabular}{cccccc}
\hline
ID & $\dot{M}/\dot{M}_{\rm Edd}$ & $L/L_{\rm Edd}$ & $r_{\mathrm{in}}/r_{g}$ & Time interval   & $\rho_0$ \\
   &                             &                 &                         & ($r_g/c$)       & (${\rm g\ cm}^{-3}$)         \\
\hline
1  & $0.35 \pm 0.09$             & $0.3 \pm 0.08$  & $22 \pm 5$              & $35815 -40930$  & $1.5 \times 10^{-6}$ \\ 
2  & $0.24 \pm 0.07$             & $0.2 \pm 0.07$  & $26 \pm 5$              & $35815 -40930$  & $1 \times 10^{-6}$ \\ 
3  & $0.11 \pm 0.02$             & $0.09 \pm 0.02$& $41 \pm 10$              & $35815 -40930$  & $5 \times 10^{-7}$ \\ 
4  & $0.06 \pm 0.01$             & $0.046\pm0.008$ & $55 \pm 12$             & $32400-37520$   & $3 \times 10^{-7}$ \\ 
5  & $0.02 $                     & $0.015\pm0.008$ & $>100$                  & $32400-37520$   & $1 \times 10^{-7}$ \\
\hline
\end{tabular}
\caption{Time-averaged mass accretion rates, Eddington ratios and thin disk inner radii for the numerical simulations. The ID column indicates the number assigned to each model. Column 5 shows the time intervals over which the averages were computed. Column 6 shows the maximum torus density in CGS units. }
\label{tab}
\end{table*}

\begin{figure*}
\centering
\includegraphics[width=\linewidth]{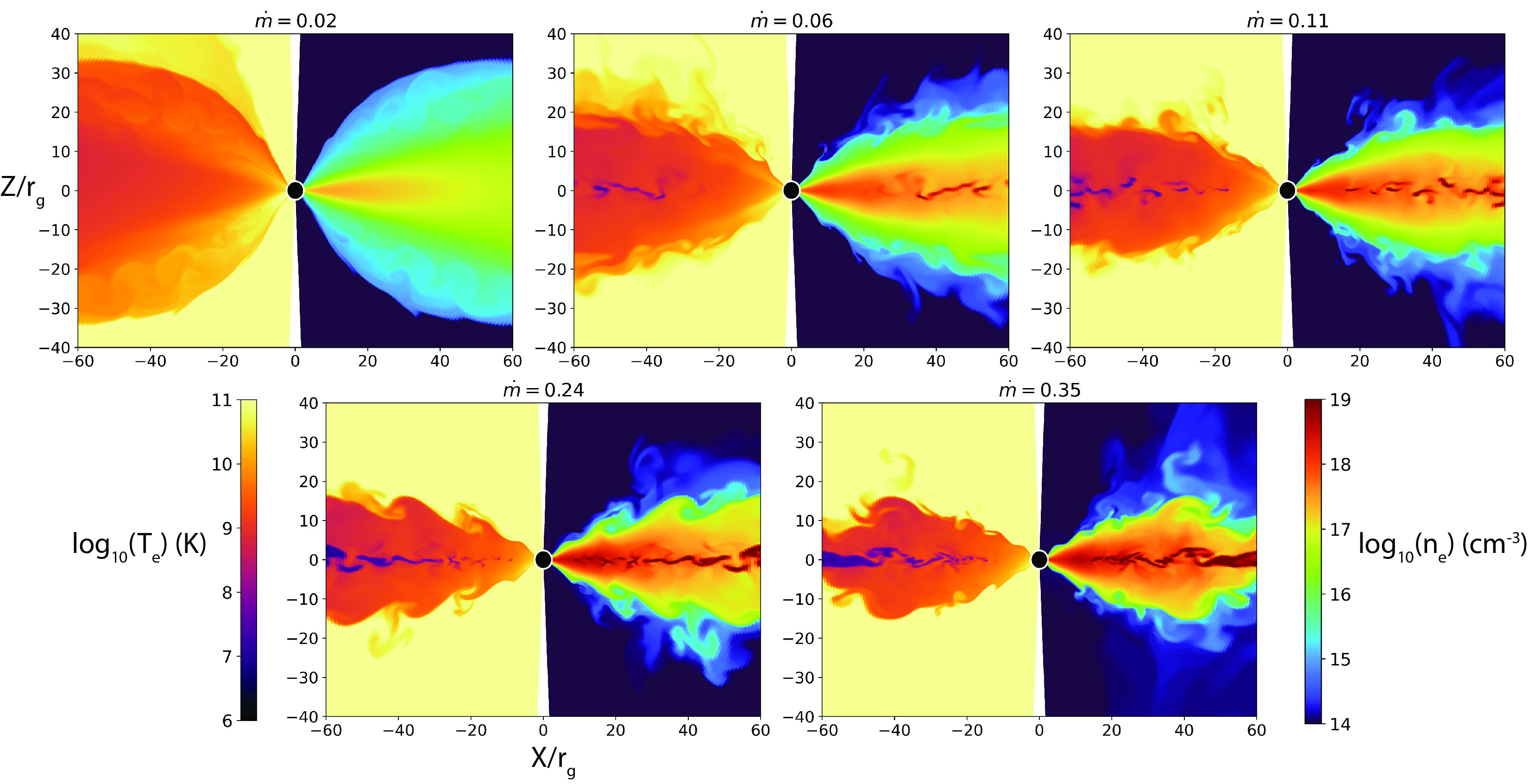}
\caption{Temperature and density maps of the five simulations carried out in this work. On the left side, the colors indicate the log of the electron temperature in Kelvin; on the right side, they indicate the log of electron number density in particles per cubic cm. All snapshots were taken at $t = 34112GM/c^3$. }
\label{fig:sims}
\end{figure*}

A snapshot from the simulations can be seen in Figure \ref{fig:sims} where the electron density and temperature maps are shown. 
Our simulations demonstrate that for $\dot{m} \geq 0.06$ the initial torus collapses resulting in the formation of a cold thin disk embedded in a hot corona, as required to explain XRBs. The dense disk is located at the equatorial plane and is 10-100 times denser than the torus in non-radiative simulations. The corona in our models typically has $T_e \sim 10^{9-10} K$ while the disk has typical electron temperatures $\sim 10^7 K$. 

Over the duration of the simulation, the cold and hot phases coexist until the accretion rate is decreased to $\dot{m} = 0.02$. At this value, we could not find any trace of a cold disk. The only structure remaining is a geometrically thick, low-density, hot corona. This is expected because as the accretion rate gets lower the density also decreases, and the gas starts becoming inefficient in radiating i.e. the cooling time is longer than the viscous time. 

We have also found that the coronal size and the inner radius of the thin disk change noticeably across the different simulations, in response to the $\dot{m}$ variation. We estimate the time-averaged coronal scale height $H_c$ and the inner radius $R_{\rm in}$ using the same time range adopted to compute $\dot{m}$ (Table \ref{tab}). The uncertainties in these estimates reflect the standard deviation in the time-averages. For more details on the methods for obtaining $H_c$ and $R_{\rm in}$, please refer to Appendices \ref{ap:corona} and \ref{ap:trunc}, respectively.

Figure \ref{fig:corona} shows that the corona contracts as the accretion rate increases. Quantitatively, as $\dot{m}$ increases sixfold from 0.06 to 0.35 the corona contracts to about sixty per cent of the size at the lowest $\dot{m}$ simulated. This is reflected in both its lateral width and height. We fitted a linear relation to the dependence of $H_c$ on $\dot{m}$ in log-log space and found that it is approximately given by 
\begin{equation}
H_c = 20 r_g \left( \frac{\dot{m}}{0.1} \right)^{-0.37 \pm 0.08}.
\end{equation}
Figure \ref{fig:corona} also displays the inner radius of the thin disk as a function of the mass accretion rate. We find that the inner radius of the thin disk is anticorrelated with $\dot{m}$. In other words, as the accretion rate increases, the disk gets closer to the event horizon. We fitted a power-law to our data and obtained 
\begin{equation}
R_{\rm in} = 42 r_g \left( \frac{\dot{m}}{0.1} \right)^{-0.60 \pm 0.27}.
\end{equation}

\begin{figure}
\begin{center}
\includegraphics[width=0.95 \columnwidth]{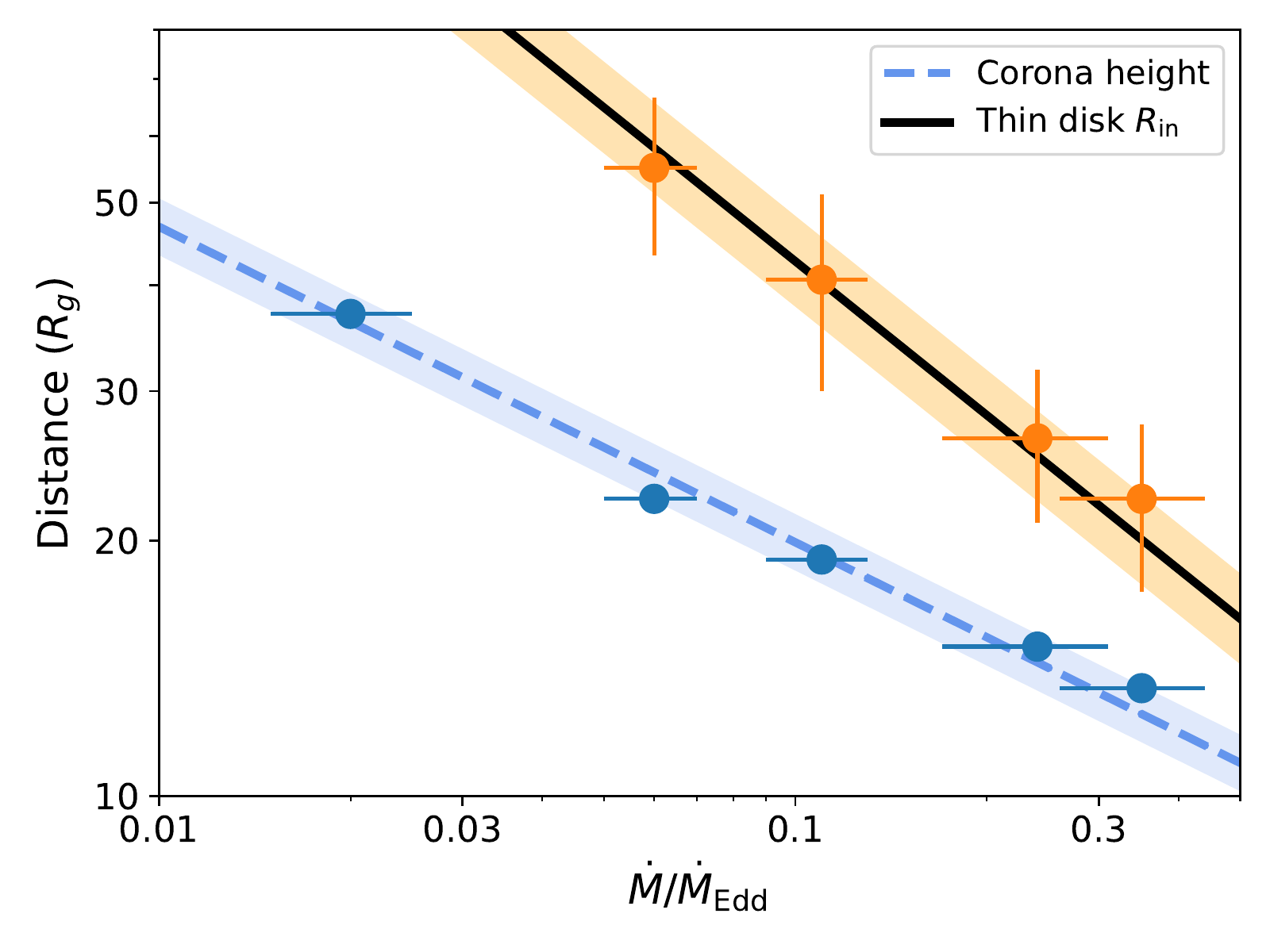}
\caption{Corona vertical height and inner radius of the thin disk as a function of the mass accretion rate $\dot{m}$. The corona and inner radius contract as $\dot{m}$ increases. Each line indicate a linear fit in log-log space using $y|x$ BCES regression with bootstrapping taking into account uncertainties in both axes \eg{Nemmen2012}. The band around each fit denotes the $1\sigma$ prediction band. }
\label{fig:corona}
\end{center}
\end{figure}

\section{Discussion}    \label{sec:disc}

Our simulations begin with a hot torus which collapses to a colder thin disk due to thermal instability. A hot corona persists above and below the thin disk. The corona and the inner radius of the thin disk gradually increase in size as $\dot{m}$ decreases. At $\dot{m}=0.02$ the thin disk disappears and there is only a hot accretion flow left. Thus, we constrain the critical accretion rate at which the thin disk disappears entirely to be in the range $0.02<\dot{m}_{\text {crit }} \lesssim 0.06$. This interval is consistent with estimates based on analytical theory which find values in the range $0.01\lesssim \dot{m}_{\text {crit }} \lesssim 0.1$ \citep{Narayan1995, Xie2012}. In one of their radiation general relativistic magnetohydrodynamic (GRRMHD) simulations (model \code{M3}), \cite{Dexter2021} find tentative collapse of the hot accretion flow at $\dot{m} \sim 10^{-3}$; this is one order of magnitude lower than our $\dot{m}_{\text {crit}}$.

\subsection{Comparison with observations}

It is worth comparing our values of $R_{\rm in}$ with measurements from observations of X-ray binaries. Since observers quote the source luminosity while our simulations are parameterized in terms of $\dot{m}$, we estimate $L$ from $\dot{m}$ for our models using equation $L=\eta\dot{M}c^2$ with the efficiency of conversion of accreted rest-mass energy into radiation given by $\eta \equiv \eta_{\rm RIAF}+\eta_{\rm thin}$. $\eta_{\rm RIAF}=\eta_{\rm RIAF}(\dot{m})$ is based on the RIAF model of \cite{Xie2012} assuming an electron turbulent heating efficiency of $\delta=0.1$; $\eta_{\rm thin}=\eta_{\rm thin}(R_{\rm in}, R_{\rm out})$ corresponds to a truncated thin disk given by equation 4 of \cite{Reb2018} where we fix $R_{\rm out}=10^4 r_g$. 

\begin{figure*}
\begin{center}
\includegraphics[width=\linewidth]{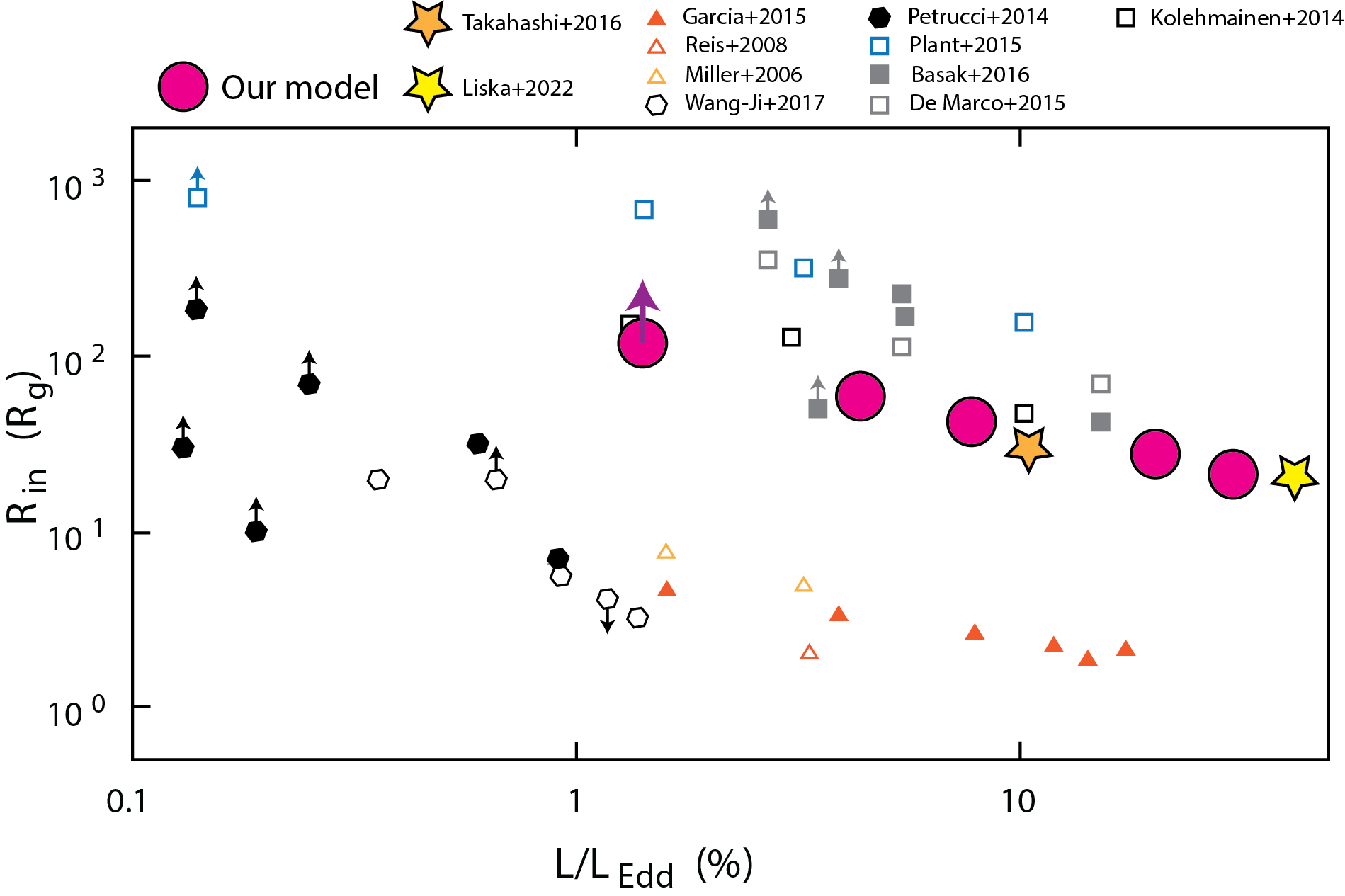}
\caption{Inner radius of the thin accretion disk as a function of the luminosity in Eddington units. $R_{\rm in}$ is in units of the gravitational radius. Circles indicate the results from our numerical simulations. Squares, triangles and hexagons denote measurements based on observations of X-ray binary GX 339-4 in the hard state \citep{Reis2008, Miller2006, Petrucci2014, Kolehmainen2014, Garcia2015, Plant2015, Basak2016, De-Marco2015, Wang-Ji2018}. The stars show results from the simulations of \protect\cite{Takahashi2016} and \protect\cite{Liska2022}. }
\label{fig:Rin}
\end{center}
\end{figure*}

Since it is widely agreed that in the thermal state the disk is truncated at or near the innermost stable circular orbit radius, here we compare our results with XRBs in the hard state. Figure \ref{fig:Rin} compares the $R_{\rm in}$ values as a function of $L/L_{\rm Edd}$ from our simulations with observations of GX 339-4 in the hard state. Figure \ref{fig:Rin} illustrates the disk truncation controversy: the inferred inner edge of the disk varies widely among different techniques even for the same object. One of the methods for measuring $R_{\rm in}$ is based on fitting the disk reflection component observed with spectroscopy \eg{Garcia2015}. Two other methods are fitting the continuum spectrum of the accretion flow \citep{McClintock2014} and timing analysis of X-ray lags \eg{De-Marco2015}. The observations form two separate tracks in Figure \ref{fig:Rin}: the upper track is roughly consistent with $R_{\rm in} \sim 74 r_g ( \lambda/0.1 )^{-0.9 \pm 0.2}$ ($\lambda \equiv L/L_{\rm Edd}$) while the lower track follows $R_{\rm in} \sim 2 r_g ( \lambda/0.1 )^{-0.6 \pm 0.1}$. 

From our simulations we obtain $R_{\rm in} \approx 40 r_g ( \lambda/0.1 )^{-0.5}$. In other words, the simulations are naturally predicting a radius-luminosity trend consistent with observations---namely, $R_{\rm in}$ approaches the BH as the luminosity increases. While the characteristic simulated $R_{\rm in}$ values are closer to the upper track, the slope of the radius-luminosity correlation is consistent the lower track. One common aspect of all the studies predicting the largest truncation (i.e. the upper track) is that the observations were taken with XMM-Newton’s Epic PN camera in timing mode, which is prone to calibration issues and pile-up effects at the very high count rates reached by the source and could potentially bias their results \citep{Garcia2015}. 

Our results indicate that with increasing Eddington ratios the corona contracts but does not go away, even at luminosities typical of the soft state ($0.1 < L/L_{\rm Edd} < 0.4$; cf. Table \ref{tab}). One possibility is that the compact corona seen in our models at such luminosities could be responsible for the hard X-ray tail observed in the soft state of several sources \eg{Sridhar2020, Connors2021}. More work is needed to probe the stability properties and electromagnetic spectrum of this remnant corona.

\subsection{Caveats}

The specific values of $R_{\rm in}$ and $L/L_{\rm Edd}$ from the models should be taken with care due to the many shortcomings of our numerical simulations described below. 

We compute the synchrotron cooling rate based on a single frequency---the critical frequency $\nu_c$ of self-absorbed synchrotron emission; $\nu_c$ is also used as the seed photon frequency for comptonization; comptonization is assumed to happen only locally. These simplifications were adopted in order to reduce the computational costs and as a result we could be underpredicting the amount of radiative cooling. 

Since we performed only hydrodynamical simulations, the dynamical effect of magnetic fields is only incorporated via the assumption of an isotropically-tangled magnetic field in equipartition with the gas, with a constant $\beta$. MHD simulations show that it is not constant as assumed here \eg{Hirose2004}. Furthermore, viscous stresses are incorporated using an $\alpha$-viscosity prescription proportional to the density. This means that most of the dissipation takes place at the equatorial plane which concentrates most of the flowing mass. MHD simulations indicate that the vertical distribution of energy dissipation can extend to higher heights \eg{Huang2023}. 

At accretion rates $\dot{m} \gtrsim 0.1$, radiation pressure becomes important and is not incorporated in this work. In addition, gravity in our model is pseudo-Newtonian. This is very good approximation for the Schwarzschild spacetime. However, the innermost stable circular orbit is closer to the horizon in the Kerr spacetime for $a_* > 0$ ($a_*$ is the BH spin) compared to the Schwarzschild spacetime. Indeed, the ISCO around a BH spinning at $a_*=0.998$ is about five times smaller compared to a nonrotating hole. We suspect that the Schwarzschild spacetime introduces a positive bias in the values of $R_{\rm in}$, which could explain why our values are systematically larger than the measured ones. 

Our coronal temperatures reach $10^{9-10} K$, hotter than XRB corona measurements \eg{Garcia2015, Chakraborty2020}, suggesting we might be missing important physics related to cooling. All the above effects can impact the results, particularly the corona size, thin disk inner radius, their dependence on $\dot{m}$ and the $\dot{m}_{\rm crit}$.

XRBs display a hysteresis loop in the hardness-luminosity diagram in which they move from the hard to soft state along a higher luminosity than the soft to hard \eg{Remillard2006}. This would correspond to a loop in the $R_{\rm in}-L/L_{\rm Edd}$ diagram, which our simulations are presently unable to account for. We think additional physics is required beyond the dependence of cooling on $\dot{m}$. Different causes have been proposed for the observed hysteresis, including the interplay between advected magnetic fields and turbulence \citep{Begelman2014}, warped disks \citep{Nixon2014}, the relation between $\alpha$ and the magnetic Prandtl number \citep{Balbus2008} and the disk magnetization \citep{Petrucci2008}.

Our simulations have a duration of at most $t_{\rm sim} \approx 4 \times 10^4 r_g/c$. In order to achieve inflow equilibrium, $t_{\rm sim}$ would need to be of the order of the viscous time for the thin disk \eg{Frank2002} which scaled to the relevant XRB parameters becomes 
\begin{align}
t_{\rm visc} (M, \dot{m}, r) = 3 \times 10^6 \frac{r_g}{c} \left( \frac{\alpha}{0.1} \right)^{-0.8} \left( \frac{M}{10 M_\odot} \right)^{1.2}  \times \nonumber \\
\left( \frac{r}{10 r_g} \right)^{1.25}  \left( \frac{\dot{m}}{0.1} \right)^{-0.3}.
\end{align}
At the radius where most of action is happening in our models, $r \approx 50 r_g$, $t_{\rm visc}$ is about $8 \times 10^6 r_g/c$ (or 15 minutes in physical units). Therefore, our models do not achieve inflow equilibrium. We would need to evolve them for durations at least a hundred times longer in order to achieve a steady state solution. For this reason, it is difficult to ascertain the robustness of our measured truncation radii. 
As far as we know, no global simulation in the literature currently has a long enough duration to ensure that thin accretion disks achieve inflow equilibrium. Future simulations---with much longer durations to reach $t_{\rm sim} \sim t_{\rm visc}$ and correspondingly larger tori such that the BH does not run out of gas---are warranted.

\subsection{Comparison with other simulations}

We now compare our work with other numerical simulations of truncated accretion disks. \cite{Hogg2017} performed MHD simulations of the transition state of XRBs with a $-1/r$ gravitational potential. Their initial condition is a thin disk with an ad-hoc cooling function. Due to their choice of cooling function, their simulation gives $R_{\rm in} \approx 90 r_g$ at $\dot{m} = 0.01$ by construction. Other numerical works have improved on many of the issues we pointed previously. \cite{Takahashi2016} performed GRRMHD simulations of truncated discs around black holes including radiation pressure in the M1 closure. For $\dot{m} \approx 0.1$, Takahashi et al. find $R_{\rm in} \approx 30 r_g$ which is consistent with our results within uncertainties. 
\cite{Dexter2021} performed GRRMHD simulations of a MAD flow onto a Kerr BH with frequency-dependent Monte Carlo radiative transfer, including synchrotron and bremsstrahlung emission. In one of their simulations, M3, they reported the tentative formation of a thin disc at $\dot{m} \sim 10^{-3}$. Dexter et al. did not report whether their collapsed disk is truncated at the inner edge. Finally, \cite{Liska2022} simulated a XRB with a GRRMHD formalism including electron thermodynamics and M1 closure, with an initial condition broadly similar to ours. They also find a two-phase flow at $\dot{m} = 0.2$ with the thin disk truncated at $R_{\rm in} \approx 20 r_g$. Liska et al. argue that the inner edge of their radiation pressure-supported thin disk is dictated by the onset of a magnetically dominated flow. 

The comparison of our results with other numerical simulations suggests that the formation of a thin disk sandwiched by a hot corona is a surprisingly robust result. Furthermore, despite their vastly different complexity, the GRRMHD models of \cite{Takahashi2016,Liska2022} produce truncated thin disks whose inner edge locations are consistent with the results of our models within $1\sigma$ uncertainties (cf. Figure \ref{fig:Rin}). The cause of truncation, however, differs among the simulations. In ours---which does not include magnetic fields---it is thermal instability; in Liska et al. it is due to a magnetic fields. There are different routes to truncation; it remains to be seen which one is taken by nature.

\section{Summary and conclusions}   \label{sec:end}

We performed two-dimensional hydrodynamical simulations of two-temperature accretion flows onto a $10 M_\odot$ black hole, incorporating radiative cooling from bremmstrahlung, synchrotron and comptonized synchrotron to energy losses. Schwarzschild spacetime is mimicked using a pseudo-Newtonian potential. Our motivation was to investigate the hard state of X-ray binaries and the hard-to-soft state transition. As such, we varied the mass accretion rate in the range $0.02 \leq \dot{M}/\dot{M}_{\rm Edd} \leq 0.35$ near the critical value at which a RIAF is thought to collapse into a radiatively efficient disk. Our main results are:

(i) Our simulations result in the formation of a cold thin disk embedded in a hot corona, as required to explain the hard state of XRBs. The typical electron temperatures are $10^{9-10} K$ for the corona and $10^{7} K$ for the thin disk. 

(ii) By varying the mass accretion rate in our simulations, we find that the corona contracts as $\dot{m}$ increases following the relation $H_c = 20 r_g ( \dot{m}/0.1 )^{-0.37}$ where $H_c$ is the corona scale height and $\dot{m}$ is in Eddington units.

(iii) The inner edge of the thin disk approaches the event horizon as $\dot{m}$ increases following the relation $R_{\rm in} = 42 r_g \left( \dot{m}/0.1 \right)^{-0.60}$.

(iv) At $\dot{m}=0.02$ the thin disk disappears and there is only a hot accretion flow left. We constrain the critical accretion rate at which the thin disk disappears entirely to be in the range $0.02<\dot{m}_{\text {crit }} \lesssim 0.06$.

These new results are important because they shed light on the details of the hard state and the hard-to-soft state transition. When a stellar BH emerges from quiescence it has a hard X-ray spectrum produced by a centrally located and compact hot corona, which then changes to a soft spectrum dominated by the thin disk emission \eg{Done2007, Fender2012}. Much debate persists over how this transition occurs. In particular, it is not clear whether the transition is driven by a reduction in the inner radius of the disk or by a shrinking corona \citep{Plant2014, Fabian2014, Garcia2015}. For example, the X-ray observations of MAXI J1820+070 indicate that the culprit is the reduction in the spatial extent of the corona \citep{Kara2019}. Contrary to simple expectations, however, our models suggest that \textit{the truncation radius decreases and the corona shrinks as the XRB evolves during the hard state raising}.

We find that the thin disk is embedded in the hot corona rather than the latter being above the disk as in a simple ``lamp post'' picture \citep{Done2007}. Our values of $R_{\mathrm{in}}$---much like those obtained in current GRRMHD models---are systematically larger than many reflection spectroscopy XRB measurements. As BH accretion models become increasingly more sophisticated, it remains to be seen whether this discrepancy between theory and observations will be resolved, and which of the two mechanisms---thermal or magnetic---are responsible for the disk truncation.

\section*{Data Availability}

Please refer to \url{https://github.com/black-hole-group/pluto}, branch \code{cooling}, commit \code{ce495cc} for the specific code and radiative cooling implementations used in this work. The initial condition files are available at \url{https://bitbucket.org/nemmen/pluto-simulations} in the directory \code{hd/torus\_2D\_pn\_cool}. The simulation binary data files, cooling table and analysis notebooks used to generate the figures in this work are available on Figshare at \url{http://tiny.cc/xrb-hydro-figshare}. 

\section*{Acknowledgements}

We acknowledge useful discussions with Roger Blandford, Dan Wilkins, Elisabete de Gouveia Dal Pino, Raimundo Lopes de Oliveira Filho, Sergio Martin Alvarez, Chris Done and Andrzej Zdziarski. We are grateful to the anonymous referee for the insightful comments and suggestions, which enhanced the quality of this paper. This work was supported by the following agencies: FAPESP (Funda\c{c}\~ao de Amparo \`a Pesquisa do Estado de S\~ao Paulo) through grants 2017/25710-1, 2017/01461-2, 2019/10054-7, 2021/01563-5, 2022/07456-9 and 2022/10460-8; NASA through the Fermi Guest Investigator Program (Cycle 16). Research developed with the help of HPC resources provided by the Information Technology Superintendence of the University of S\~ao Paulo. IA acknowledge funding from an United Kingdom Research and Innovation grant (code: MR/V022830/1). The Black Hole Group used GPUs generously donated by NVIDIA under the GPU Grant Program.

\bibliographystyle{mnras}
\bibliography{refs,stanford} 

\begin{thebibliography}{}
\makeatletter
\relax
\def\mn@urlcharsother{\let\do\@makeother \do\$\do\&\do\#\do\^\do\_\do\%\do\~}
\def\mn@doi{\begingroup\mn@urlcharsother \@ifnextchar [ {\mn@doi@}
  {\mn@doi@[]}}
\def\mn@doi@[#1]#2{\def\@tempa{#1}\ifx\@tempa\@empty \href
  {http://dx.doi.org/#2} {doi:#2}\else \href {http://dx.doi.org/#2} {#1}\fi
  \endgroup}
\def\mn@eprint#1#2{\mn@eprint@#1:#2::\@nil}
\def\mn@eprint@arXiv#1{\href {http://arxiv.org/abs/#1} {{\tt arXiv:#1}}}
\def\mn@eprint@dblp#1{\href {http://dblp.uni-trier.de/rec/bibtex/#1.xml}
  {dblp:#1}}
\def\mn@eprint@#1:#2:#3:#4\@nil{\def\@tempa {#1}\def\@tempb {#2}\def\@tempc
  {#3}\ifx \@tempc \@empty \let \@tempc \@tempb \let \@tempb \@tempa \fi \ifx
  \@tempb \@empty \def\@tempb {arXiv}\fi \@ifundefined
  {mn@eprint@\@tempb}{\@tempb:\@tempc}{\expandafter \expandafter \csname
  mn@eprint@\@tempb\endcsname \expandafter{\@tempc}}}

\bibitem[\protect\citeauthoryear{{Balbus} \& {Henri}}{{Balbus} \&
  {Henri}}{2008}]{Balbus2008}
{Balbus} S.~A.,  {Henri} P.,  2008, \mn@doi [\apj] {10.1086/524838}, \href
  {https://ui.adsabs.harvard.edu/abs/2008ApJ...674..408B} {674, 408}

\bibitem[\protect\citeauthoryear{{Basak} \& {Zdziarski}}{{Basak} \&
  {Zdziarski}}{2016}]{Basak2016}
{Basak} R.,  {Zdziarski} A.~A.,  2016, \mn@doi [\mnras] {10.1093/mnras/stw420},
  \href {https://ui.adsabs.harvard.edu/abs/2016MNRAS.458.2199B} {458, 2199}

\bibitem[\protect\citeauthoryear{{Begelman}}{{Begelman}}{2014}]{Begelman2014}
{Begelman} M.~C.,  2014, preprint, \href
  {http://adsabs.harvard.edu/abs/2014arXiv1410.8132B} {} (\mn@eprint {arXiv}
  {1410.8132})

\bibitem[\protect\citeauthoryear{Chakraborty, Navale, Ratheesh  \&
  Bhattacharyya}{Chakraborty et~al.}{2020}]{Chakraborty2020}
Chakraborty S.,  Navale N.,  Ratheesh A.,   Bhattacharyya S.,  2020, \mn@doi
  [\mnras] {10.1093/mnras/staa2711}, 498, 5873

\bibitem[\protect\citeauthoryear{{Chen}, {Abramowicz}, {Lasota}, {Narayan}  \&
  {Yi}}{{Chen} et~al.}{1995}]{Chen1995}
{Chen} X.,  {Abramowicz} M.~A.,  {Lasota} J.-P.,  {Narayan} R.,   {Yi} I.,
  1995, \mn@doi [\apjl] {10.1086/187836}, \href
  {https://ui.adsabs.harvard.edu/abs/1995ApJ...443L..61C} {443, L61}

\bibitem[\protect\citeauthoryear{{Connors} et~al.,}{{Connors}
  et~al.}{2021}]{Connors2021}
{Connors} R. M.~T.,  et~al., 2021, \mn@doi [\apj] {10.3847/1538-4357/abdd2c},
  \href {https://ui.adsabs.harvard.edu/abs/2021ApJ...909..146C} {909, 146}

\bibitem[\protect\citeauthoryear{{De Marco}, {Ponti}, {Mu{\~n}oz-Darias}  \&
  {Nandra}}{{De Marco} et~al.}{2015}]{De-Marco2015}
{De Marco} B.,  {Ponti} G.,  {Mu{\~n}oz-Darias} T.,   {Nandra} K.,  2015,
  \mn@doi [\apj] {10.1088/0004-637X/814/1/50}, \href
  {https://ui.adsabs.harvard.edu/abs/2015ApJ...814...50D} {814, 50}

\bibitem[\protect\citeauthoryear{{Dermer}, {Liang}  \& {Canfield}}{{Dermer}
  et~al.}{1991}]{Dermer1991}
{Dermer} C.~D.,  {Liang} E.~P.,   {Canfield} E.,  1991, \mn@doi [\apj]
  {10.1086/169770}, \href
  {https://ui.adsabs.harvard.edu/abs/1991ApJ...369..410D} {369, 410}

\bibitem[\protect\citeauthoryear{{Dexter}, {Scepi}  \& {Begelman}}{{Dexter}
  et~al.}{2021}]{Dexter2021}
{Dexter} J.,  {Scepi} N.,   {Begelman} M.~C.,  2021, \mn@doi [\apjl]
  {10.3847/2041-8213/ac2608}, \href
  {https://ui.adsabs.harvard.edu/abs/2021ApJ...919L..20D} {919, L20}

\bibitem[\protect\citeauthoryear{{Done}, {Gierli{\'n}ski}  \& {Kubota}}{{Done}
  et~al.}{2007}]{Done2007}
{Done} C.,  {Gierli{\'n}ski} M.,   {Kubota} A.,  2007, \mn@doi [\aapr]
  {10.1007/s00159-007-0006-1}, \href
  {http://adsabs.harvard.edu/abs/2007A%26ARv..15....1D} {15, 1}

\bibitem[\protect\citeauthoryear{{Esin}, {McClintock}  \& {Narayan}}{{Esin}
  et~al.}{1997}]{Esin1997}
{Esin} A.~A.,  {McClintock} J.~E.,   {Narayan} R.,  1997, \mn@doi [\apj]
  {10.1086/304829}, \href {http://adsabs.harvard.edu/abs/1997ApJ...489..865E}
  {489, 865}

\bibitem[\protect\citeauthoryear{{Fabian}, {Parker}, {Wilkins}, {Miller},
  {Kara}, {Reynolds}  \& {Dauser}}{{Fabian} et~al.}{2014}]{Fabian2014}
{Fabian} A.~C.,  {Parker} M.~L.,  {Wilkins} D.~R.,  {Miller} J.~M.,  {Kara} E.,
   {Reynolds} C.~S.,   {Dauser} T.,  2014, \mn@doi [\mnras]
  {10.1093/mnras/stu045}, \href
  {https://ui.adsabs.harvard.edu/abs/2014MNRAS.439.2307F} {439, 2307}

\bibitem[\protect\citeauthoryear{{Fender} \& {Belloni}}{{Fender} \&
  {Belloni}}{2012}]{Fender2012}
{Fender} R.,  {Belloni} T.,  2012, \mn@doi [Science] {10.1126/science.1221790},
  \href {http://adsabs.harvard.edu/abs/2012Sci...337..540F} {337, 540}

\bibitem[\protect\citeauthoryear{{Frank}, {King}  \& {Raine}}{{Frank}
  et~al.}{2002}]{Frank2002}
{Frank} J.,  {King} A.,   {Raine} D.~J.,  2002, {Accretion Power in
  Astrophysics}, 3rd edn.
Cambridge University Press

\bibitem[\protect\citeauthoryear{{Garc{\'\i}a}, {Steiner}, {McClintock},
  {Remillard}, {Grinberg}  \& {Dauser}}{{Garc{\'\i}a}
  et~al.}{2015}]{Garcia2015}
{Garc{\'\i}a} J.~A.,  {Steiner} J.~F.,  {McClintock} J.~E.,  {Remillard} R.~A.,
   {Grinberg} V.,   {Dauser} T.,  2015, \mn@doi [\apj]
  {10.1088/0004-637X/813/2/84}, \href
  {https://ui.adsabs.harvard.edu/abs/2015ApJ...813...84G} {813, 84}

\bibitem[\protect\citeauthoryear{{Hirose}, {Krolik}, {De Villiers}  \&
  {Hawley}}{{Hirose} et~al.}{2004}]{Hirose2004}
{Hirose} S.,  {Krolik} J.~H.,  {De Villiers} J.-P.,   {Hawley} J.~F.,  2004,
  \mn@doi [\apj] {10.1086/383184}, \href
  {http://adsabs.harvard.edu/abs/2004ApJ...606.1083H} {606, 1083}

\bibitem[\protect\citeauthoryear{Hogg \& Reynolds}{Hogg \&
  Reynolds}{2017}]{Hogg2017}
Hogg J.~D.,  Reynolds C.~S.,  2017, \apj, 843, 80

\bibitem[\protect\citeauthoryear{Huang, Jiang, Feng, Davis, Stone  \&
  Middleton}{Huang et~al.}{2023}]{Huang2023}
Huang J.,  Jiang Y.-F.,  Feng H.,  Davis S.~W.,  Stone J.~M.,   Middleton
  M.~J.,  2023, \mn@doi [\apj] {10.3847/1538-4357/acb6fc}, 945, 57

\bibitem[\protect\citeauthoryear{{Hubeny}}{{Hubeny}}{1990}]{Hubeny1990}
{Hubeny} I.,  1990, \mn@doi [\apj] {10.1086/168501}, \href
  {https://ui.adsabs.harvard.edu/abs/1990ApJ...351..632H} {351, 632}

\bibitem[\protect\citeauthoryear{{Ingram} \& {Done}}{{Ingram} \&
  {Done}}{2011}]{Ingram2011}
{Ingram} A.,  {Done} C.,  2011, \mn@doi [\mnras]
  {10.1111/j.1365-2966.2011.18860.x}, \href
  {https://ui.adsabs.harvard.edu/abs/2011MNRAS.415.2323I} {415, 2323}

\bibitem[\protect\citeauthoryear{Kara et~al.,}{Kara et~al.}{2019}]{Kara2019}
Kara E.,  et~al., 2019, Nature, 565, 198

\bibitem[\protect\citeauthoryear{{Kolehmainen}, {Done}  \& {D{\'\i}az
  Trigo}}{{Kolehmainen} et~al.}{2014}]{Kolehmainen2014}
{Kolehmainen} M.,  {Done} C.,   {D{\'\i}az Trigo} M.,  2014, \mn@doi [\mnras]
  {10.1093/mnras/stt1886}, \href
  {https://ui.adsabs.harvard.edu/abs/2014MNRAS.437..316K} {437, 316}

\bibitem[\protect\citeauthoryear{{Landau} \& {Lifshitz}}{{Landau} \&
  {Lifshitz}}{1959}]{Landau1959}
{Landau} L.~D.,  {Lifshitz} E.~M.,  1959, {Fluid mechanics}

\bibitem[\protect\citeauthoryear{{Liska}, {Musoke}, {Tchekhovskoy}, {Porth}  \&
  {Beloborodov}}{{Liska} et~al.}{2022}]{Liska2022}
{Liska} M.~T.~P.,  {Musoke} G.,  {Tchekhovskoy} A.,  {Porth} O.,
  {Beloborodov} A.~M.,  2022, \mn@doi [\apjl] {10.3847/2041-8213/ac84db}, \href
  {https://ui.adsabs.harvard.edu/abs/2022ApJ...935L...1L} {935, L1}

\bibitem[\protect\citeauthoryear{McClintock, Narayan  \& Steiner}{McClintock
  et~al.}{2014}]{McClintock2014}
McClintock J.~E.,  Narayan R.,   Steiner J.~F.,  2014, \mn@doi [Space Science
  Reviews] {10.1007/s11214-013-0003-9}, 183, 295

\bibitem[\protect\citeauthoryear{{Mignone}, {Bodo}, {Massaglia}, {Matsakos},
  {Tesileanu}, {Zanni}  \& {Ferrari}}{{Mignone} et~al.}{2007}]{Mignone2007}
{Mignone} A.,  {Bodo} G.,  {Massaglia} S.,  {Matsakos} T.,  {Tesileanu} O.,
  {Zanni} C.,   {Ferrari} A.,  2007, \mn@doi [\apjs] {10.1086/513316}, \href
  {http://adsabs.harvard.edu/abs/2007ApJS..170..228M} {170, 228}

\bibitem[\protect\citeauthoryear{{Miller}, {Homan}, {Steeghs}, {Rupen},
  {Hunstead}, {Wijnands}, {Charles}  \& {Fabian}}{{Miller}
  et~al.}{2006}]{Miller2006}
{Miller} J.~M.,  {Homan} J.,  {Steeghs} D.,  {Rupen} M.,  {Hunstead} R.~W.,
  {Wijnands} R.,  {Charles} P.~A.,   {Fabian} A.~C.,  2006, \mn@doi [\apj]
  {10.1086/508644}, \href
  {https://ui.adsabs.harvard.edu/abs/2006ApJ...653..525M} {653, 525}

\bibitem[\protect\citeauthoryear{{Narayan} \& {Yi}}{{Narayan} \&
  {Yi}}{1995a}]{Narayan1995b}
{Narayan} R.,  {Yi} I.,  1995a, \mn@doi [\apj] {10.1086/175599}, \href
  {http://adsabs.harvard.edu/abs/1995ApJ...444..231N} {444, 231}

\bibitem[\protect\citeauthoryear{{Narayan} \& {Yi}}{{Narayan} \&
  {Yi}}{1995b}]{Narayan1995}
{Narayan} R.,  {Yi} I.,  1995b, \mn@doi [\apj] {10.1086/176343}, \href
  {http://adsabs.harvard.edu/abs/1995ApJ...452..710N} {452, 710}

\bibitem[\protect\citeauthoryear{{Nemmen}, {Georganopoulos}, {Guiriec},
  {Meyer}, {Gehrels}  \& {Sambruna}}{{Nemmen} et~al.}{2012}]{Nemmen2012}
{Nemmen} R.~S.,  {Georganopoulos} M.,  {Guiriec} S.,  {Meyer} E.~T.,  {Gehrels}
  N.,   {Sambruna} R.~M.,  2012, \mn@doi [Science] {10.1126/science.1227416},
  \href {http://adsabs.harvard.edu/abs/2012Sci...338.1445N} {338, 1445}

\bibitem[\protect\citeauthoryear{{Nixon} \& {Salvesen}}{{Nixon} \&
  {Salvesen}}{2014}]{Nixon2014}
{Nixon} C.,  {Salvesen} G.,  2014, \mn@doi [\mnras] {10.1093/mnras/stt2215},
  \href {http://adsabs.harvard.edu/abs/2014MNRAS.437.3994N} {437, 3994}

\bibitem[\protect\citeauthoryear{{Pacholczyk}}{{Pacholczyk}}{1970}]{Pacholczyk1970}
{Pacholczyk} A.~G.,  1970, {Radio astrophysics. Nonthermal processes in
  galactic and extragalactic sources}

\bibitem[\protect\citeauthoryear{{Paczy{\'n}sky} \& {Wiita}}{{Paczy{\'n}sky} \&
  {Wiita}}{1980}]{Paczynsky1980}
{Paczy{\'n}sky} B.,  {Wiita} P.~J.,  1980, \aap, \href
  {http://adsabs.harvard.edu/abs/1980A%26A....88...23P} {88, 23}

\bibitem[\protect\citeauthoryear{{Papaloizou} \& {Pringle}}{{Papaloizou} \&
  {Pringle}}{1984}]{Papaloizou1984}
{Papaloizou} J.~C.~B.,  {Pringle} J.~E.,  1984, \mn@doi [\mnras]
  {10.1093/mnras/208.4.721}, \href
  {http://adsabs.harvard.edu/abs/1984MNRAS.208..721P} {208, 721}

\bibitem[\protect\citeauthoryear{Petrucci, Ferreira, Henri  \&
  Pelletier}{Petrucci et~al.}{2008}]{Petrucci2008}
Petrucci P.-O.,  Ferreira J.,  Henri G.,   Pelletier G.,  2008, \mn@doi
  [\mnras] {10.1111/j.1745-3933.2008.00439.x}, 385, L88

\bibitem[\protect\citeauthoryear{{Petrucci}, {Cabanac}, {Corbel}, {Koerding}
  \& {Fender}}{{Petrucci} et~al.}{2014}]{Petrucci2014}
{Petrucci} P.~O.,  {Cabanac} C.,  {Corbel} S.,  {Koerding} E.,   {Fender} R.,
  2014, \mn@doi [\aap] {10.1051/0004-6361/201322268}, \href
  {https://ui.adsabs.harvard.edu/abs/2014A&A...564A..37P} {564, A37}

\bibitem[\protect\citeauthoryear{{Piran}, {S{\k{a}}dowski}  \&
  {Tchekhovskoy}}{{Piran} et~al.}{2015}]{Piran2015}
{Piran} T.,  {S{\k{a}}dowski} A.,   {Tchekhovskoy} A.,  2015, \mn@doi [\mnras]
  {10.1093/mnras/stv1547}, \href
  {https://ui.adsabs.harvard.edu/abs/2015MNRAS.453..157P} {453, 157}

\bibitem[\protect\citeauthoryear{{Plant}, {Fender}, {Ponti}, {Mu{\~n}oz-Darias}
   \& {Coriat}}{{Plant} et~al.}{2014}]{Plant2014}
{Plant} D.~S.,  {Fender} R.~P.,  {Ponti} G.,  {Mu{\~n}oz-Darias} T.,   {Coriat}
  M.,  2014, \mn@doi [\mnras] {10.1093/mnras/stu867}, \href
  {https://ui.adsabs.harvard.edu/abs/2014MNRAS.442.1767P} {442, 1767}

\bibitem[\protect\citeauthoryear{{Plant}, {Fender}, {Ponti}, {Mu{\~n}oz-Darias}
   \& {Coriat}}{{Plant} et~al.}{2015}]{Plant2015}
{Plant} D.~S.,  {Fender} R.~P.,  {Ponti} G.,  {Mu{\~n}oz-Darias} T.,   {Coriat}
  M.,  2015, \mn@doi [\aap] {10.1051/0004-6361/201423925}, \href
  {https://ui.adsabs.harvard.edu/abs/2015A&A...573A.120P} {573, A120}

\bibitem[\protect\citeauthoryear{{Reb}, {Fern{\'a}ndez-Ontiveros}, {Prieto}  \&
  {Dolag}}{{Reb} et~al.}{2018}]{Reb2018}
{Reb} L.,  {Fern{\'a}ndez-Ontiveros} J.~A.,  {Prieto} M.~A.,   {Dolag} K.,
  2018, \mn@doi [\mnras] {10.1093/mnrasl/sly079}, \href
  {https://ui.adsabs.harvard.edu/abs/2018MNRAS.478L.122R} {478, L122}

\bibitem[\protect\citeauthoryear{{Reis}, {Fabian}, {Ross}, {Miniutti}, {Miller}
   \& {Reynolds}}{{Reis} et~al.}{2008}]{Reis2008}
{Reis} R.~C.,  {Fabian} A.~C.,  {Ross} R.~R.,  {Miniutti} G.,  {Miller} J.~M.,
   {Reynolds} C.,  2008, \mn@doi [\mnras] {10.1111/j.1365-2966.2008.13358.x},
  \href {https://ui.adsabs.harvard.edu/abs/2008MNRAS.387.1489R} {387, 1489}

\bibitem[\protect\citeauthoryear{{Remillard} \& {McClintock}}{{Remillard} \&
  {McClintock}}{2006}]{Remillard2006}
{Remillard} R.~A.,  {McClintock} J.~E.,  2006, \mn@doi [\araa]
  {10.1146/annurev.astro.44.051905.092532}, \href
  {http://adsabs.harvard.edu/abs/2006ARA%26A..44...49R} {44, 49}

\bibitem[\protect\citeauthoryear{{Schnittman}, {Krolik}  \&
  {Noble}}{{Schnittman} et~al.}{2013}]{Schnittman2013}
{Schnittman} J.~D.,  {Krolik} J.~H.,   {Noble} S.~C.,  2013, \mn@doi [\apj]
  {10.1088/0004-637X/769/2/156}, \href
  {http://adsabs.harvard.edu/abs/2013ApJ...769..156S} {769, 156}

\bibitem[\protect\citeauthoryear{{Shakura} \& {Sunyaev}}{{Shakura} \&
  {Sunyaev}}{1973}]{Shakura1973}
{Shakura} N.~I.,  {Sunyaev} R.~A.,  1973, \aap, \href
  {http://adsabs.harvard.edu/abs/1973A%26A....24..337S} {24, 337}

\bibitem[\protect\citeauthoryear{{Sridhar}, {Garc{\'\i}a}, {Steiner},
  {Connors}, {Grinberg}  \& {Harrison}}{{Sridhar} et~al.}{2020}]{Sridhar2020}
{Sridhar} N.,  {Garc{\'\i}a} J.~A.,  {Steiner} J.~F.,  {Connors} R. M.~T.,
  {Grinberg} V.,   {Harrison} F.~A.,  2020, \mn@doi [\apj]
  {10.3847/1538-4357/ab64f5}, \href
  {https://ui.adsabs.harvard.edu/abs/2020ApJ...890...53S} {890, 53}

\bibitem[\protect\citeauthoryear{{Stepney} \& {Guilbert}}{{Stepney} \&
  {Guilbert}}{1983}]{Stepney1983}
{Stepney} S.,  {Guilbert} P.~W.,  1983, \mn@doi [\mnras]
  {10.1093/mnras/204.4.1269}, \href
  {https://ui.adsabs.harvard.edu/abs/1983MNRAS.204.1269S} {204, 1269}

\bibitem[\protect\citeauthoryear{{Stone}, {Pringle}  \& {Begelman}}{{Stone}
  et~al.}{1999}]{Stone1999}
{Stone} J.~M.,  {Pringle} J.~E.,   {Begelman} M.~C.,  1999, \mn@doi [\mnras]
  {10.1046/j.1365-8711.1999.03024.x}, \href
  {http://adsabs.harvard.edu/abs/1999MNRAS.310.1002S} {310, 1002}

\bibitem[\protect\citeauthoryear{Takahashi, Ohsuga, Kawashima  \&
  Sekiguchi}{Takahashi et~al.}{2016}]{Takahashi2016}
Takahashi H.~R.,  Ohsuga K.,  Kawashima T.,   Sekiguchi Y.,  2016, \apj, 826,
  23

\bibitem[\protect\citeauthoryear{Vemado}{Vemado}{2020}]{Vemado2020}
Vemado A.,  2020, Master's thesis, Universidade de Sao Paulo,
  https://doi.org/10.11606/D.14.2020.tde-14052020-172856

\bibitem[\protect\citeauthoryear{Wang-Ji et~al.,}{Wang-Ji
  et~al.}{2018}]{Wang-Ji2018}
Wang-Ji J.,  et~al., 2018, \mn@doi [The Astrophysical Journal]
  {10.3847/1538-4357/aaa974}, 855, 61

\bibitem[\protect\citeauthoryear{{Wang} et~al.,}{{Wang}
  et~al.}{2021}]{Wang2021}
{Wang} J.,  et~al., 2021, \mn@doi [\apjl] {10.3847/2041-8213/abec79}, \href
  {https://ui.adsabs.harvard.edu/abs/2021ApJ...910L...3W} {910, L3}

\bibitem[\protect\citeauthoryear{{Wu}, {Xie}, {Yuan}  \& {Gan}}{{Wu}
  et~al.}{2016}]{Wu2016}
{Wu} M.-C.,  {Xie} F.-G.,  {Yuan} Y.-F.,   {Gan} Z.,  2016, \mn@doi [\mnras]
  {10.1093/mnras/stw742}, \href
  {http://adsabs.harvard.edu/abs/2016MNRAS.459.1543W} {459, 1543}

\bibitem[\protect\citeauthoryear{{Xie} \& {Yuan}}{{Xie} \&
  {Yuan}}{2012}]{Xie2012}
{Xie} F.-G.,  {Yuan} F.,  2012, \mn@doi [\mnras]
  {10.1111/j.1365-2966.2012.22030.x}, \href
  {http://adsabs.harvard.edu/abs/2012MNRAS.427.1580X} {427, 1580}

\bibitem[\protect\citeauthoryear{{Yuan} \& {Narayan}}{{Yuan} \&
  {Narayan}}{2014}]{Yuan2014}
{Yuan} F.,  {Narayan} R.,  2014, \mn@doi [\araa]
  {10.1146/annurev-astro-082812-141003}, \href
  {http://adsabs.harvard.edu/abs/2014ARA%26A..52..529Y} {52, 529}

\makeatother
\end{thebibliography}

\appendix

\section{Heating and cooling}   \label{ap:energy}

Here, we describe technical details of the heating and cooling implementation. Two source terms need to be added to the equations in order to incorporate (i) heating and angular momentum removal due to magnetically-driven turbulent viscosity and (ii) energy losses due to radiative cooling. 

We assume an $\alpha$-disk \citep{Shakura1973} and adopt the ``K-model'' viscosity prescription in \cite{Stone1999}, which corresponds to the kinematic viscosity $\nu = \alpha r^{1/2}$ where $\alpha$ is a constant similar to (but not equal) the standard Shakura-Sunyaev $\alpha$. We set $\alpha = 0.01$ which is a common choice \eg{Wu2016}.

The radiative cooling rate is given by 
\begin{equation}
Q^{-}_{\rm rad}=Q^{-}_{\rm brem} + Q^{-}_{\rm syn} + Q^{-}_{\rm ssc}
\label{Qrad}
\end{equation}
where we include the relevant processes for stellar-mass BHs: bremsstrahlung emission $Q^{-}_{\rm brem}$ including both the electron-ion and electron-electron interactions, optically thin synchrotron emission $Q^{-}_{\rm syn}$ and comptonized self-synchrotron emission $Q^{-}_{\rm ssc}$. We adopt a series of approximations for the calculation of the cooling rates, as follows.

\subsection{Bremsstrahlung}

The bremsstrahlung emission is given by 
\begin{equation}
Q^{-}_{\rm brem}=Q^{-}_{ei} + Q^{-}_{ee}
\end{equation}
where the subscript $ei$ and $ee$ denote the electron-ion and the electron-electron rates. In this approach, we assume that the proton temperature is small enough to neglect its motion and the electrons follow a Maxwellian energy distribution. The spectral emissivity is integrated and the fit gives \citep{Stepney1983}
\begin{equation}\label{Qei}
Q^{-}_{ei}=1.48\times 10^{-22}n_e^2 \times F(\theta_e) \ {\rm erg \ cm}^{-3} {\rm s}^{-1},
\end{equation}
where $F(\theta_e)$ is given by different expressions depending on the value of $\theta_e \equiv k T_e/m_e c^2$:
\begin{equation}
F(\theta_e) =
\begin{cases}
\frac{9\theta_e}{2\pi}\left[\ln\left(1.123\theta_e + 0.48\right) + 1.5\right] & \text{if } \theta_e < 1,\\ 
4\left(\frac{2\theta_e}{\pi^3}\right)^{1/2}\left(1+1.781\theta_e^{1.34}\right) & \text{if } \theta_e > 1.
\end{cases}
\end{equation}

For the electron-electron rate,  \citet{Stepney1983} gives the photon spectrum and integrate it with an approximated cross-section, resulting in  
\begin{equation}
Q^{-}_{ee} =
\begin{cases}
2.56\times 10^{-22}n_e^2 \theta_e^{3/2} 
\left(1+1.1\theta_e+\theta_e^2-1.25\theta_e^{5/2}\right) & \text{if } \theta_e < 1,\\ 
3.40\times 10^{-22}n_e^2 \theta_e \left(\ln{1.123\theta_e}+1.28\right) & \text{if } \theta_e > 1,
\end{cases}
\end{equation}
in units of ${\rm erg \ cm}^{-3} {\rm s}^{-1}$. With these approximations, the formulas above incur in errors of less than 5 per cent \citep{Stepney1983}. Some of the constants here are different from \cite{Stepney1983} to ensure smoothness across $\theta_e = 1$ \citep{Narayan1995}.

\subsection{Synchrotron}

For the synchrotron emission, we adopt the optically thin limit with a relativistic Maxwellian distribution of electrons \citep{Pacholczyk1970}. We assume that the angle between the electron velocity vector and the direction of the local magnetic field is isotropically distributed; this incurs an error of less than 2 per cent \citep{Narayan1995}. We estimate the self-absorption critical frequency $\nu_c$ by assuming that at each radius $r$, the synchrotron emission occurs over a volume $2 H \Delta s$ where $H$ is the vertical scale height ($H = R c_s /v_K$), $c_s$ is the sound speed, $v_K$ is the Keplerian velocity and $\Delta s$ is the surface area following \cite{Wu2016}. Matching the synchrotron and Rayleigh-Jeans blackbody emission results in a transcendental equation for $x_m$, 
\begin{align} \label{transeq}
\exp{\left(1.8899 x_m^{1/3}\right)}=2.49\times 10^{-10}\frac{12\pi n_e H}{B} \frac{1}{\theta_e^3 K_2(1/\theta_e)} \times \\ 
\left(\frac{1}{x_m^{7/6}}+\frac{0.40}{x_m^{17/12}}+\frac{0.5316}{x_m^{5/3}}\right),
\end{align}
where $K_2$ is the modified Bessel's function. This equation is solved at each $H$ (i.e. each $r$) to obtain $x_m$. It is then used to calculate $\nu_c$ from 
\begin{equation} 
\nu_c = \frac{3}{2} \nu_o \theta_e^2 x_m
\end{equation}
where $\nu_o = 2.8\times10^6 B$ Hz with $B$ is Gauss.

We obtain the cooling rate per unit volume by matching the total cooling over a shell with the net flux reaching an observer at infinity and assuming that at each frequency $\nu$ the observer sees a blackbody with radius determined by the condition $\nu = \nu_c(r)$. This results in
\begin{equation}    \label{Qsyn}
Q^{-}_{\rm syn} = \frac{2\pi}{3c^2}k T_e(r)\frac{d\nu_c^3(r)}{dr} \approx \frac{2\pi}{3c^2}k T_e(r)\frac{\nu_c^3(r)}{r}.
\end{equation}
Equation \ref{Qsyn} is written in a form that explicitly ensures that the integral of $Q^{-}_{\rm syn}$ over the entire flow is equal to the total cooling radiation reaching infinity.

\subsection{Comptonization}

We neglect nonlocal radiative transfer effects, i.e. photons being generated in one place and scattered in another. The Compton cooling is mainly characterized by the Comptonized energy enhancement factor $\eta$. This parameter is defined as the average change in energy of a photon between injection and escape. An approximate description for $\eta$ is given by \citet{Dermer1991} as
\begin{equation}    \label{eta}
\eta = 1 + \frac{P(A-1)}{1-PA}\left[1-\left(\frac{x}{3\theta_e}\right)^{-1-\frac{\ln{P}}{\ln{A}}}\right]
\end{equation}
where $x \equiv h\nu/(m_e c^2)$, $P$ is the probability of photon scattering, 
\begin{equation}    \label{P}
P = 1 - e^{-\tau_{\rm s}},
\end{equation}
$\tau_{\rm s}=2 n_e \sigma_T H$ is the scattering optical depth and $A$ is the mean amplification factor in the energy of a scattered photon, considering a Maxwellian velocity distribution of temperature $\theta_e$ for the electrons,
\begin{equation}    \label{A}
A = 1 + 4\theta_e + 16\theta_e^2.
\end{equation}

In the synchrotron cooling model, the escaping radiation is emitted mostly at the local self-absorption cutoff frequency $\nu_c$. The Comptonization of this radiation gives an additional cooling rate of
\begin{equation}\label{Qssc}
Q^{-}_{\rm ssc} = Q^{-}_{\rm syn}\left[\eta(\nu_c, \tau_{s}) - 1\right].
\end{equation}

\subsection{Optically thick regime}

Since we have the formation of dense regions in our simulations, it is important to compute the radiative cooling of the higher density, colder, optically thick regions of the accretion flow associated with the thin disk. We adopt the net cooling rate of \citet{Hubeny1990}, 
\begin{equation}    \label{Qradtot}
Q^{-}_{\rm rad} = \frac{4\sigma T_e^4}{H}\frac{1}{3\tau/2 + \sqrt{3} + 1/\tau_{\rm abs}}.
\end{equation}
Here, $\tau = \tau_{\rm abs} + \tau_{s}$ is the total optical depth and $\tau_{\rm abs}$ can be approximated as 
\begin{equation}    \label{tauabs}
\tau_{\rm abs} = \frac{H}{4\sigma T_e^4} \left(Q^{-}_{\rm brem} + Q^{-}_{\rm syn} + Q^{-}_{\rm ssc}\right).
\end{equation}
(cf. equation 3.33 from \citealt{Narayan1995}). Note that this expression is valid for both optically thick and thin limits, because when $\tau \gg 1$, equation \ref{Qradtot} results in the appropriate blackbody limit. On the other hand, when $\tau \ll 1$, equation \ref{Qradtot} becomes equation \ref{Qrad}.

\section{Corona thickness}  \label{ap:corona}

Before quantifying the coronal vertical thickness $H_c$, we first convert the array with the numerical values of density from the native polar coordinates adopted in \code{PLUTO} to cylindrical coordinates $R-z$---a procedure we call regridding for which GPU-accelerated routines written by the group are used for faster calculations\footnote{The routines we developed for analyzing \code{PLUTO} dump files are available at \url{https://bitbucket.org/nemmen/mickey/}.}. The density values are then obtained as a function of $R$. 

At each value of $R$, the distribution of density as function of $z$ is a superposition of two Gaussians centered at $z=0$, the disk midplane: one with low amplitude and small width corresponding to the thin disk (if it forms), and a second broader Gaussian with large amplitude describing the RIAF. 

We compute the time-averaged density map for each model within the time intervals displayed in Table \ref{tab}. Given the time-averaged density map $\bar{\rho}_t=\rho(R,z)$, for the points where the condition $\log{(\rho/\rho_0)} \geq -0.5$ is valid the density is replaced by the threshold value, i.e. $\log \rho \rightarrow -0.5$. This procedure is necessary because as the thin disk is much denser than the corona, the distribution of densities of the thin disk will dominate in the estimates of vertical thickness. Therefore, the denser part corresponding to the thin disk must be removed. One example of thin disk removal is displayed in Figure \ref{fig:corona}. 

\begin{figure*}
\begin{center}
\includegraphics[width=0.8\textwidth]{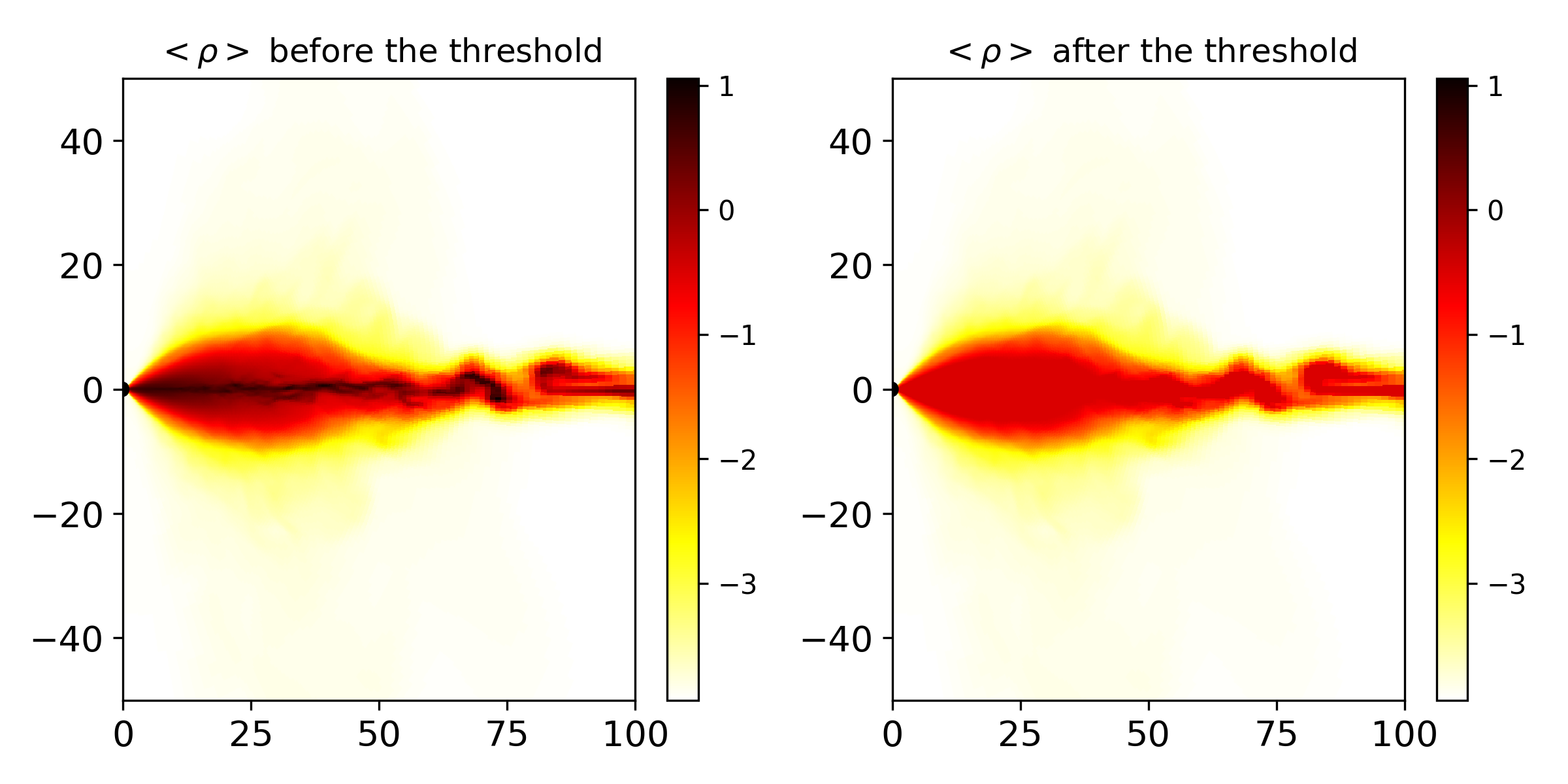}
\caption{Result of removing the thin disk for estimating the corona properties. Here, the time-averaged density of model 3 is displayed.} 
\label{fig:corona}
\end{center}
\end{figure*}

At each value of $R$ we fit a Gaussian function to the $\rho(z)$ distribution, obtaining a list of accretion flow scale heights $H'(R)= \sigma (R)$ where $\sigma$ is the Gaussian width at a given $R$. Finally, we estimate the coronal thickness as $H_c = 2 \max [H'(R)] $.

\section{Inner radius of thin disk}   \label{ap:trunc}

\begin{figure}
\begin{center}
\includegraphics[width=\columnwidth,trim=0 0 0 25,clip=true]{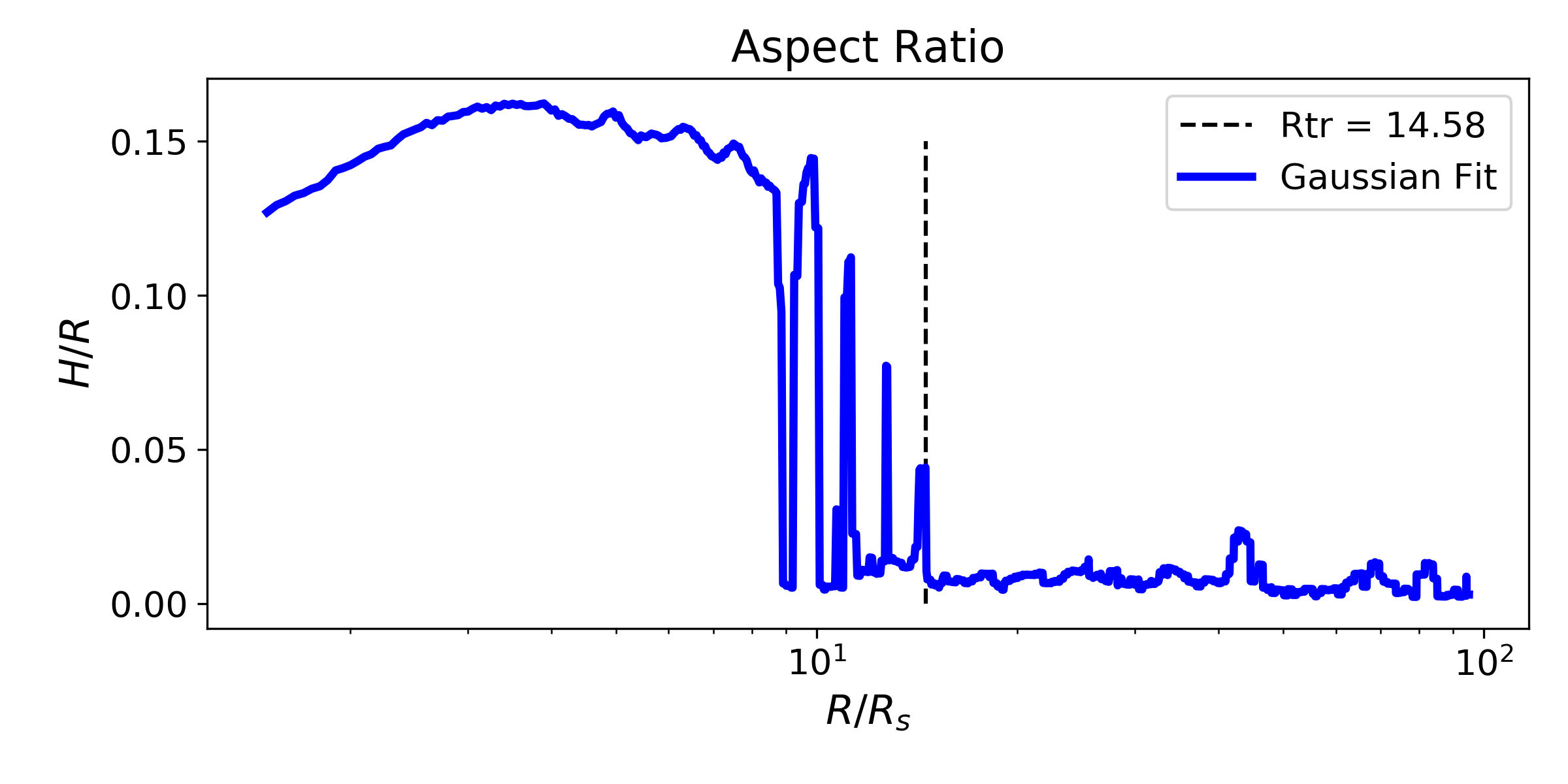}
\caption{An example of the dependency of the disk thickness $H'/R$ on the cylindrical radius $R/R_{\mathrm{s}}$, where $R_s$ is the Schwarzschild radius. The dashed line indicates the resulting inner radius based on the method outlined in the text.}
\label{fig:H/R}
\end{center}
\end{figure}

In order to estimate the inner radius of the thin disk $R_{\rm in}$, we perform the same coordinate transformation described in Appendix \ref{ap:corona} and obtain the spacetime density distribution $\rho=\rho(R,z;t)$. At each value of $R$, we fit a Gaussian function in order to characterize the vertical density distribution. The Gaussian width corresponds to the accretion flow scale height, $H'=\sigma$---not to be confused with the coronal thickness $H_c$. At each simulation time step, this gives us $H'/R$ as a function of $R$ (Figure \ref{fig:H/R}). 

When the flow collapses into a thin disk, the density increases by a factor of $10-100$ times the RIAF equatorial density. The Gaussian width in this case corresponds to the vertical density distribution of the thin disk instead of the RIAF. We define the thin accretion disk as the cold and dense gas with $H'/R < 0.015$. This is consistent with the range $10^{-3} \lesssim H/R \lesssim 0.01$ expected in accretion disk theory $10^{-3}-0.01$ \citep{Piran2015, Frank2002}. The thin disks in our simulations settle to vertical equilibrium with $H'/R > 4 \times 10^{-3}$. Our results are not very sensitive to the particular value of the threshold thickness as long as it is not much larger than $0.01$.

We notice in Figure \ref{fig:H/R} that the transition between the hot accretion flow ($H'/R \approx 0.15$) and the cold thin disk ($H'/R \lesssim 0.01$) is not sharp. We define $R_{\rm in}$ as the value of $R$ at which the condition $H'/R < 0.015$ is attained based on the moving average computed using 80 points starting from smaller $R$. This definition ensures that the inner radius is located where the oscillations in the value of $H'/R$ reaches a minimum (cf. dashed line in Figure \ref{fig:H/R}). 

\begin{figure*}
\begin{minipage}{.48\textwidth}
  \centering
  \includegraphics[width=\linewidth]{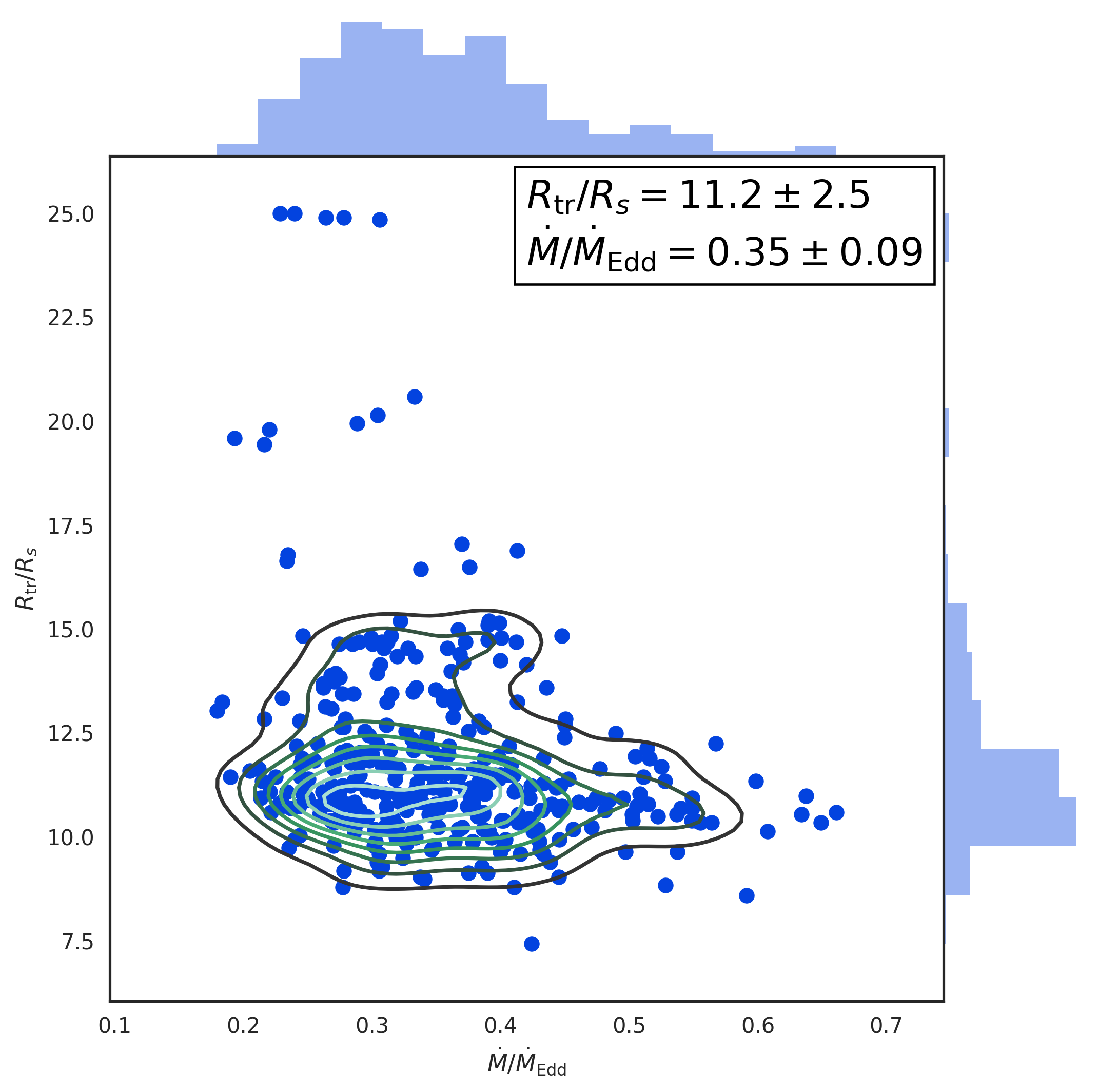}
\end{minipage}
\begin{minipage}{.48\textwidth}
  \centering
  \includegraphics[width=\linewidth]{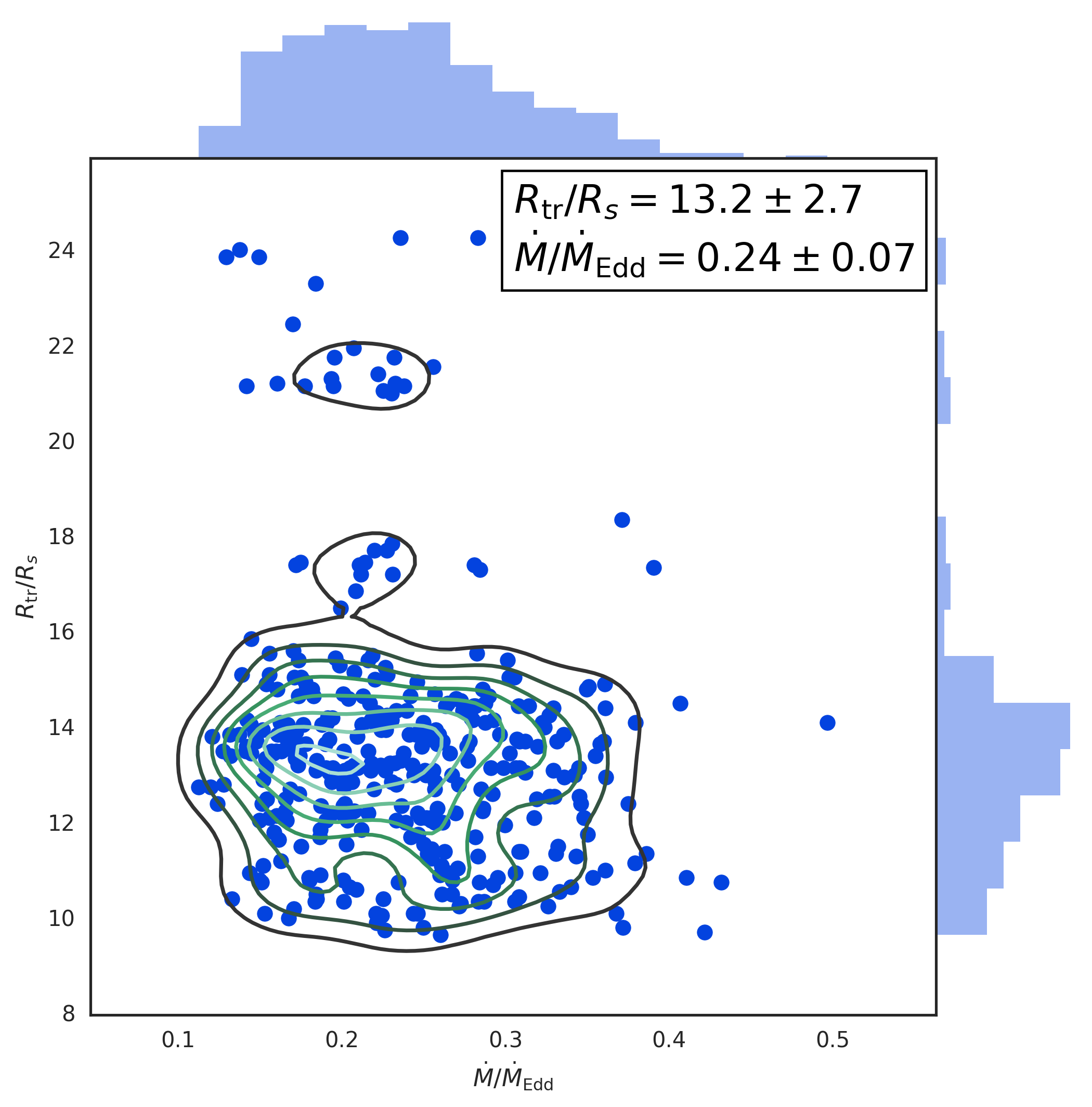}
\end{minipage}
\begin{minipage}{.48\textwidth}
  \centering
  \includegraphics[width=\linewidth]{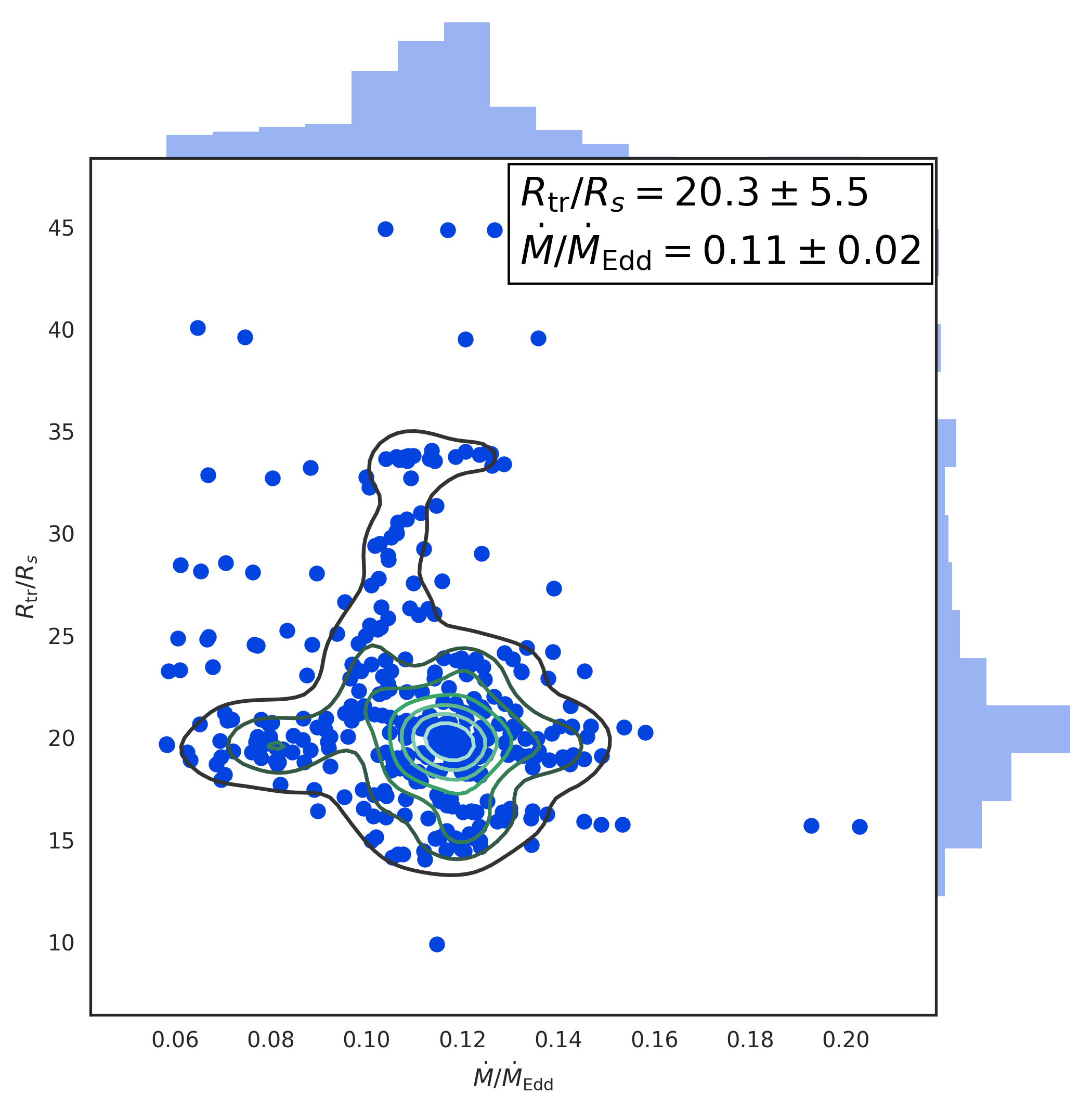}
\end{minipage}
\begin{minipage}{.48\textwidth}
  \centering
  \includegraphics[width=\linewidth]{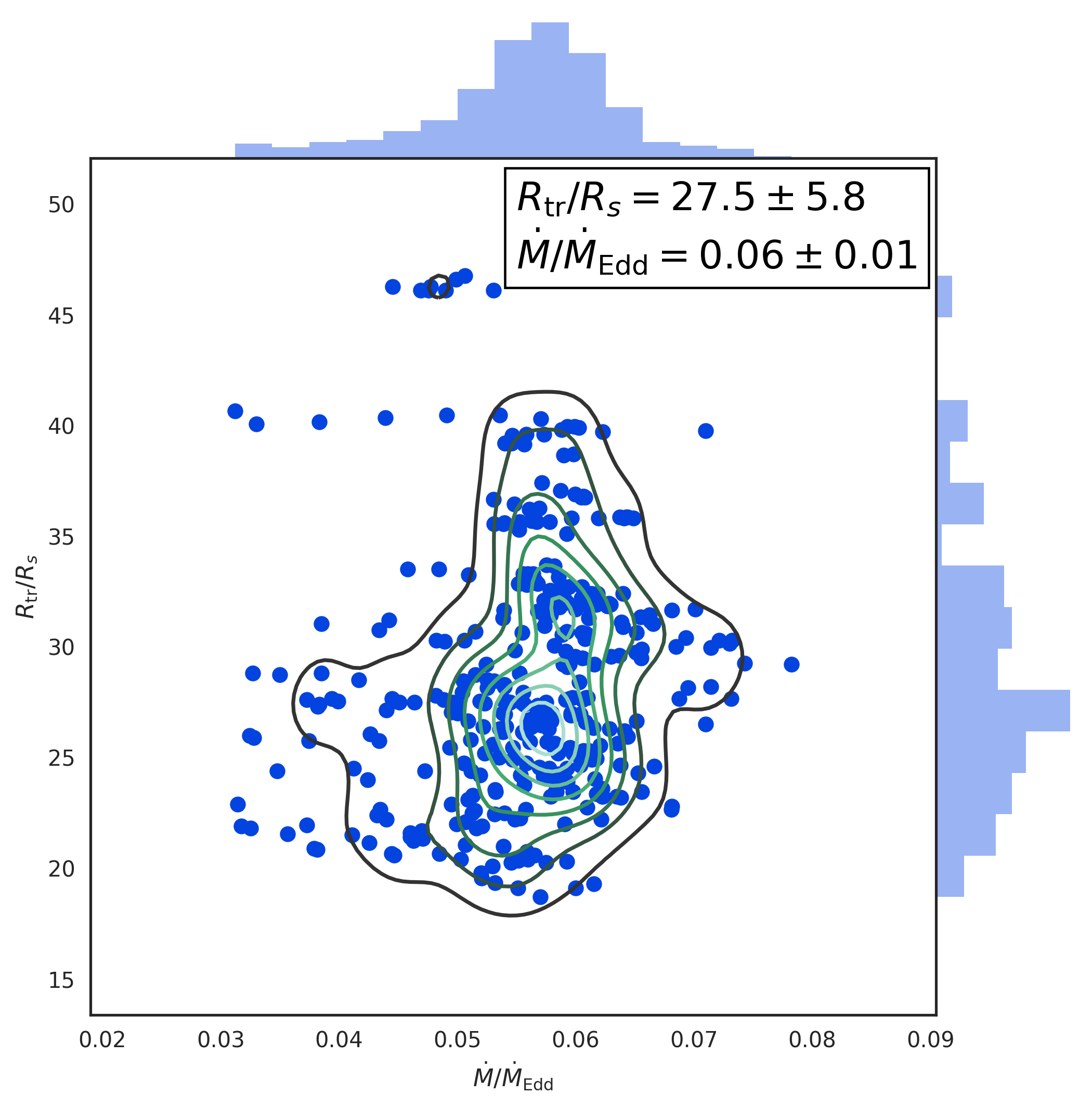}
\end{minipage}
\caption{Distribution of inner radii for the simulations where a thin disk emerged, as a function of the mass accretion rate in Eddington units. Each point corresponds to the value of $R_{\rm in}$ computed using the procedure outlined in Appendix \ref{ap:trunc} for a given timestep. }
\label{fig:Rin-mdot}
\end{figure*}

Figure \ref{fig:Rin-mdot} shows a scatter plot of $R_{\rm in}$ versus $\dot{m}$ for the four models where a thin disk emerged. This figure illustrates the variability of the inner radius. For a given model, the final inner radius is computed as the median of $R_{\rm in}$ using the time range indicated in Table \ref{tab}. The uncertainty that we quote in our $R_{\rm in}$ estimates correspond to the standard deviation. 

\bsp	
\label{lastpage}
\end{document}